# Modeling the stylized facts of wholesale system marginal price (SMP) and the impacts of regulatory reforms on the Greek Electricity Market

George P. Papaioannou [a], PhD,
Panagiotis G. Papaioannou [b]
Nikos Parliaris [c], MSc

[a] Center for Research and Applications of Nonlinear Systems, CRANS,
University of Patras Greece (Adjunct Researcher) and
ADMIE S.A., the Independent Power Transmission Operator, Greece.
[b] National Technical University of Athens,
Department of Applied Mathematics and Physical Sciences and ARVIUS S.A (CEO & Co-Founder)
[c] Independent Expert in Financial Markets and ARVIUS S.A (President & Co-Founder)

2013

# Abstract

This work presents the results of an empirical research with the target of modeling the stylized facts of the daily ex-post System Marginal Price (SMP) of the Greek wholesale electricity market, using data from January 2004 to December of 2011. SMP is considered here as the footprint of an underline stochastic and nonlinear process that bears all the information reflecting not only the effects of changes in endogenous or fundamental factors of the market but also the impacts of a series of regulatory reforms that have continuously changed the market's microstructure. To capture the dynamics of the conditional mean and volatility of SMP that generate the stylized facts (mean-reversion, price spikes, fat - tails price distribution etc), a number of ARMAX/GARCH models have been estimated using as regressors an extensive set of fundamental factors in the Greek electricity market, as well as dummy variables that mimic the history of Regulator's interventions. The findings show that changes in the microstructure of the market caused by the reforms have strongly affected the dynamic evolution of SMP and that the best found model captures adequately the stylized facts of the series that other electricity and financial markets share. The dynamics of the conditional volatility generated by the model can be extremely useful in the efforts that are under way towards market restructuring so the Greek market to be more compatible with the requirements of the European Target Model.

## 1. Introduction

Liberalization and deregulation of Electricity Markets, two facts that have characterize, in the last decades, this industry, are the main drivers, behind a gradual increase of the degree of structural complexity. Therefore, the decision making process of the market's agents (generators, distributors, suppliers, system operators, regulators etc.) is extremely difficult due to uncertainty, nonlinearity and unpredictability that are inherent characteristics of any high dimensional and nonlinear complex structure. The most representative feature of the complexity of electricity markets is its price **volatility** and the associated with it **stylized facts** i.e the nature of its statistical properties. Volatility is roughly the continuous change with time of the standard deviation of price changes, **conditional** on the most recent information available, a concept that is central in this paper. Electricity price volatility has a set of peculiar "footprints" (the aforementioned stylized facts), which are universal statistical characteristics commonly encountered in financial markets (Campbell J., et al, 1997, Taylor J.S., 2005).

There will be cases in which demand exceed supply (e.g when Unit availability is constrained or equivalently load is curtailed) and competitive pricing is impossible, causing **prices to exhibit high volatility** i.e. **price spikes** the level and duration of which are difficult to forecast (an ideal condition for the exercise of market power). Inadequacies also of the **Transmission system** enhance the complexity of Electricity markets which become more prone to problems with local market power

The Greek Electricity Market, GEM henceforth, as all other Day-Ahead markets, exhibits a high degree of volatility. This is because both demand-side and supply-side of the market depend on a large number of fundamental variables i.e. the phase space of the dynamics of the market is high dimensional and very complex. Price of fuel and $CO_2$, Hydropower reserves, Power Plant Unit availability, Renewable Energy Station (RES) power generation, are some of the supply variables while labor factors, other macroeconomic variables as well as temperature constitute the set of demand factors. Another factor that has increase the degree of complexity (and so the price volatility) of GEM is the gradual and dramatic increase of the number of its agents (generators, suppliers, marketers or qualified energy brokers and of licensed companies that trade electricity), during the last five years. As a consequence of this high volatility is the need for the adoption in the market of a number of hedging instruments used in financial markets to reduce the degree of uncertainty, inherent in volatile markets. This is also a requirement of the Target Model described below.

Greek Electricity Market, is a component of the Central – South European regional market (CSE Region), together with France, Italy, Germany, Austria, and Slovenia, heading towards its coupling, in the end of 2014, with the Unified European Electricity Market or Target Model[1] which is in a process of developing. To this target, RAE, the Regulator of the market, proposes its drastic restructuring in order to alleviate all possible current distortions and simultaneously to enhance the degree of competition within the market as well as its structural compatibility requirements of the target model. The most pronounced structural incompatibilities are: a) GEM's indirect auctioning system of short-term interconnection rights b) the lack of intra-day market component c) market's operation schedules not coordinated with other markets d) the co-optimization of Energy and Ancillary services e) its DAS technical solution f) Market's clearance issues g) regulated max/min Bidding prices h) variable cost recovery mechanism and j) its incompatible structure to adopt a **forward market, FM.**

The introduction of FM is by itself one of the structural changes in the market that has to be made so GEM can be well "fitted" to the Target Model. Forward Contracts are expected to be introduced in the GEM for example in the form of Generation Bids auctioning, written in such a way to account for different zones of Consumption and having different maturity time as well as contracts related to Financial or Physical Transmission Rights (RAE 2011).

---

[1] The principles of designing this new European Electricity Marker are described in details in the site of ACER (Agency for the Cooperation of Energy Regulators).

However, any efforts towards alleviating the distortions of the market, restructuring – reengineering the market to meet Target's model standards, have as a prerequisite an, in depth, understanding (qualitatively as well quantitatively) the underlying dynamics of the wholesale or system marginal price, SMP, as this key variable is the criterion for assessing the overall efficiency and effectiveness of the market. By adequately modeling the **stylized facts,** i.e. the dynamic evolution of SMP one can extract valuable information on the functioning of the market since the effects of any distortions and inefficiencies are "incorporated" in the dynamics of SMP. This in depth understanding of current and future dynamic evolution of the SMP can be achieved, by using models that are capable of replicating the underlying dynamics of SMP on which the pricing, in the near future, of forward Contracts will be based.

We provide at this point a short list of papers on stylized facts. Cont (Cont R., 2001) argues that a complete as well as representative list of stylized statistical properties of **financial asset returns** is as follows: **Absence of autocorrelations, Gain/loss asymmetry, Heavy tails, Aggregation Gaussianity, Intermittency, Volatility clustering, Conditional heavy tails, Slow decay of autocorrelation in absolute returns, Leverage effect, Volume/volatility correlation** and finally, **asymmetry in time scales**. Weron (Weron R., 2006) provides for electricity prices returns, a similar list of stylized facts that indirectly incorporates all the above, as follows: Price Spikes, Seasonality, mean-reversion and Heavy tails in the distribution of price returns. In the work of C. Dipeng and D. Bunn (Dipeng C. and D. Bunn, 2010) the stylized facts of **mean-reverting** and **spiky** characteristics of spot prices are captured by a **regime-switching** model and they are considered to be as the nonlinear effects of some exogenous factors acting on spot prices. Exogenous factors as demand, reserve margin, large fuel price changes, Carbon emission allowances and market concentration-power have nonlinear effects on spot prices and consequently on their stylized facts. Regime-switching models are also used by Deng (1998), Huisman and Mahieu (2003), Ethier and Mount (1999), to capture price spikes. Huisman (Huisman R., 2009) provides an extensive review of the characteristics or stylized facts of electricity prices. In their recent work Huisman and Kilic (Huisman R. and M. Kilic, 2011) examine the evolution of seasonality, mean reversion, time-varying volatility and price spikes in the day-ahead electricity prices for the Belgian, Dutch, German, French and Scandinavian markets over the years 2003-2010. The impact of changes in the level of reserve margin on price spike is examined by Mount (Mount T. et al, 2006), by employing a regime-switching model. It is shown that the probability of a spike increases in periods with low reserve margins. Stylized facts of the Spanish daily spot electricity prices and especially the **leverage effect** characteristic is examined (see section 7 in this work), by using a threshold asymmetric autoregressive stochastic volatility (TA-ARSV) model, are described in the work of Montero et. al (Montero J.M. et al, 2010). Huisman (Huisman R., 2008) employed switching-regimes models and found that the probability of **price spike occurrence** increases when temperature deviates substantially from its mean level. Schwartz (Schwartz E.S., 1997) has shown that commodity prices, exhibit **strong mean reversion.** Pilipovic (Pilipovic D., 1997), Weron (Weron R., and B. Przybylowicz, 2000) and Simonsen (Simonsen I., 2003) also have observed that electricity spot prices exhibit **strong mean reversion** as well. If jump behavior is included in the models, the accuracy in capturing the **stylized facts** is increased as it has been shown by Kaminsky (1997), Barz and Johnson (1998), Deng (1998), (Deng et al., 2001) and Kamat and Oren ( 2001). **Long range correlations**, measured by Hurst exponent, are examined by Erzgraber (2008). **Volatility clustering** in electricity markets and its comparison with the one exhibited by financial markets is described by Weron (Weron R., 2000), and Perello (Perello J. et al, 2007). Very high volatility expressed as a large number of **very large or extreme changes** in price values is investigated by Bystrom (Bystrom H., 2005), for Nord Pool, using **extreme value theory.** We refer also the works by Serletis and Rosenberg (Serletis A. and A.A. Rosenberg, 2009, 2007) Elder and Serletis (Elder J. and A. Serletis, 2008) and finally Kyrtsou et.al, (Kyrtsou C. et. al, 2009) in which evidence is provided for long memory of antipersistence form (detected by Hurst exponents or detrending moving average, DMA). The quantification of complex multiscale correlated behavior and the role of neglected nonlinearities in electricity markets and in energy **futures prices** (as opposed to spot prices) are also examined in these works, contributing further to the attempts to describe the degree of complexity of energy markets in general compared to the one of financial markets.

To our knowledge, this is the second empirical work addressing the impact on GEM's System Marginal Price of Regulatory Market Reports or Policy changes (the first attempt is the work of Kalantzis (Kalantzis K. et. al, 2012). However the main advantage of this work however is the use of an extensive number of fundamental factors of the Supply side of the market, as exogenous regression variables and the in depth analysis of the combined interactions of these quantitative variables with the qualitative or dummy variables representing the regulatory reforms and their effects on SMP. Previous similar works are those of Boffa (Boffa et. al, 2010), Bollino (Bollino et. al, 2008 a,b ) and Bosco (Bosco et.al, 2007). This paper enhances the relatively poor literature on modeling GEM's dynamics (the work of Theodorou P., et al 2008 is dealing with SMP dynamics but does not consider regulatory impacts). This issue is also related to the recent work of Bask and Widerberg (Bask M. et al, 2009) in which the interaction between market structure and stability and volatility of spot prices is examined. However, the work conceptually closer to ours, both on its general scope as well as on its specific research questions on how regulatory reforms and polices in the Italian market impact the dynamics of SMP, is the work of Petrella (Petrella et. al, 2012).

The main objective of this work is to answer the following questions: to what extent do typical characteristics or stylized facts of electricity spot prices encountered in other markets do exist in the Greek electricity market? Is there any inter-dependence between the gradual formation of the structure of the Greek electricity market (through its regulatory reforms that the market has undergone) and the dynamics of its SMP volatility, stability or complexity? How this interdependence can be detected via quantifying the associated stylized facts? How different are the stylized facts of an emerging electricity market as the Greek one, compared to other fully-developed markets as e.g. the neighbor more mature Italian market or the Nord Pool? Another goal is also to contribute to the growing literature on individual case studies and comparisons of different electricity market architectures and their associated stylized facts of prices, worldwide (Escribano et al., 2002, Andrianesis et al., 2011).

Therefore the authors believe that the findings in this work could be useful as follows:

a) To help in shedding light on any distortions – malfunctions and structural frictions that exist in GEM and more importantly in revealing and quantifying the effects of the Regulatory changes (by searching for example the side – effects of introducing, such mechanism as the Transitional Capacity Adequacy and Cost Recovery mechanisms), therefore contributing to the efforts of the appropriate markets' agents (Regulator, System Operator etc.) to detect those structural elements that have to be changed.
b) Indirectly, by using the findings in this paper to contribute to the efforts of re-structuring and re-engineering GEM, via a better understanding of the dynamic behavior of SMP, an information needed in the process of developing the forward market.

The rest of this paper is organized as follows. In section2 we briefly describe GEM, the time history of its architecture and the regulatory market reforms with the expected effects on SMP. Section 3 provides complete description of the data sets, their pre-processing and required tests before modelling. In section 4 we give the specification and fitting results of a number of models, the best model found and the necessary tests on its innovations to assess how successful the model is in replicating the stylized facts exhibited by SMP. In section 5 the interpretation of results is presented while in section 7, using the model, we perform in sample forecasts to replicate the impact of reforms on SMP and also compare GEM characteristics with those of other electricity and financial markets. The paper concludes with conclusions and propositions for further research.

2. Greek Electricity Market (GEM)

Greece has adopted in 2005 a pure mandatory pool for the wholesale electricity market. Its implementation was carried out in stages or transitional phases (2000 – 2005, 2005 – 2010 and 2010–today). The revised market architecture, launched in September 2010, completed in 2011 its first full year of application, has determined the **Day-Ahead (DA)**[2] SMP as the wholesale **market index** reminiscent to S&P or ASE index, as this price determines in great amounts the cash-flows of market's players. This market design encapsulate fully the all the requirement of the Grid and Market Operation Code of 2005. The design makes a clead distinction between the DA market and Balancing mechanism. The evolution of Index HHI[3] which measures the degree of openness of a market to competition has been reduced from the value of 10000 in years 2008 and 2009 to 6844 in 2010 and 5764 in 2011, an improvement of the market evolving to a more competitive state. However, GEM is far from being considered a competitive market.

**2.1 A brief description of GEM's architecture and microstructure**

Table 2.1 summarizes the time evolution of the net generation capacity for Greece. We observe a gradual decrease in the share of Lignite Units in the total mixture and an opposite behavior in the share of Natural Gas units due to constantly increasing number of investments of Independent Power Producers (IPPs). More specifically, CCGT Units during period 2004 – 2011 have increased their share about 124%. The market was completely dominated by PPC until 2004. Table 2.2 provides a comparative view of the fuel-Mix generation 2010-11 in the Interconnected system of GEM. Table 2.3 provides information on the market volume as well as peak power demand for the period 2005-2009. Market volume is defined as the traded volume of electricity and is equal to the annual demand (including the interconnection balance), i.e. to 52,365.8 MWh in 2010 and 51,872.3 in 2011. There are not yet any futures and OTC markets.

The Greek Electricity System consists of 13 Hydroelectric Stations, HES, installed in six river systems. For example, see Table 2.4, the Aheloos river system has three hydraulically coupled HES, Kremasta, Kastraki and Stratos. The total energy reserves of the three HES's dams is the times series of the regressor **Ahelenr** in table 2.3. The treatment of coupled HES gives a unified water value which can be interpreted as opportunity costs of using reservoir water at a specific time and not at an optimal later point of time. The HES production depends on the water flowing out of the Dam and water management must be done in such a way that the dam levels must remain within required range. If the water reserves in a Dam are high, HES Units must operate in order to avoid waste of energy due to overspills, while in case of low water reserves, operations must be constrained.

Among the Hydro variables the HES of the Aheloos river has the greater influence, followed by the impact of HES of the Aliakmon river (see correlations -0.58 and -0.37 respectively of SMP and energy reserves variables). For the period of October 15 to end of May each year, water is collected in the Dams while Hydro generator is practically zero. From beginning of June to mid-October Hydro must-run (for irrigation etc) and generation is taking place. Strong correlations are observed, as expected, between Kremasta Hydro-Dam level (in meters) and the corresponding Aheloos energy reserves (MWh) (0.98), and Polyfito Dam level and Aliakmon energy reserves (0.97).

---
[2] A short description of the Day-Ahead (DA) market is given in Appendix 1.

[3] HHI stands for Herfindahl-Hishman Index. If HHI = 10000 the market is a monopoly, if HHI > 5000 the market is over-concentrated, H > 1800 concentrated, for 1000 < HHI < 1800 efficiently competitive and HHI ≤ 1000 competitive.

**Table 2.1:** Installed capacity as of 2004, and 2007 - 2011 in the interconnected system

| Plant Type | 31.12.2004 Net installed capacity (MW) | % | 31.12.2007 Net installed capacity (MW) | % | 31.12.2008 Net installed capacity (MW) | % | 31.12.2009 Net installed capacity (MW) | % | 31.12.2010 Net installed capacity (MW) | % | 31.12.2011 Net installed capacity (MW) | % |
|---|---|---|---|---|---|---|---|---|---|---|---|---|
| Lignite | 4,808.00 | 43.92% | 4,808.00 | 40.50% | 4,808.10 | 38.69% | 4,808.10 | 38.49% | 4,746.00 | 33.74% | 4,496.00 | 31.30% |
| HFO (Heavy Fuel Oil) | 718.00 | 6.56% | 718.00 | 6.05% | 718.00 | 5.78% | 718.00 | 5.75% | 698.00 | 4.96% | 698.00 | 4.86% |
| CCGT (Gas Turb. Combin. Cycle) | 1,572.00 | 14.36% | 1,962.00 | 16.53% | 1,962.10 | 15.79% | 1,962.10 | 15.71% | 3,224.00 | 22.92% | 3,526.00 | 24.55% |
| Natural gas – other (Gas) (OCGT) | 487.00 | 4.45% | 487.00 | 4.10% | 486.80 | 3.92% | 486.80 | 3.90% | 487.00 | 3.46% | 487.00 | 3.39% |
| Hydro plants (large) | 3,017.00 | 27.56% | 3,017.00 | 25.41% | 3,016.50 | 24.27% | 3,016.50 | 24.15% | 3,018.00 | 21.46% | 3,017.00 | 21.00% |
| RES and small Cogeneration | 345.00 | 3.15% | 880.00 | 7.41% | 993.50 | 7.99% | 1,058.40 | 8.47% | 1,558.00 | 11.08% | 2,141.00 | 14.90% |
| Large-scale CHP | | | | | 334.00 | 2.69% | 334.00 | 2.67% | 334.00 | 2.37% | | |
| Other Cogeneration | | | | | 108.00 | 0.87% | 108.00 | 0.86% | | | | |
| **Total** | **10,947.00** | **100.00%** | **11,872.00** | **100.00%** | **12,427.00** | **100.00%** | **12,491.90** | **100.00%** | **14,065.00** | **100.00%** | **14,365.00** | **100.00%** |

**Table 2.2:** Fuel-Mix Generation 2010-11 in the Greek Interconnected System

|  | 2010 (TWh) | 2011 (TWh) | % difference |
|---|---|---|---|
| Lignite | 27.44 | 27.57 | 0.47 |
| Fuel Oil | 0.11 | 0.009 | -91.82 |
| Natural Gas | 10.36 | 14.85 | 43.34 |
| Large Hydro | 6.70 | 3.68 | -45.07 |
| RES | 2.04 | 2.53 | 24.02 |
| **Total Local Generation** | **46.6** | **48.6** | **4.29** |
| Net Imports | 5.70 | 3.23 | -43.33 |
| **Grand Total** | **52.35** | **51.87** | **-0.92** |

**Source:** ADMIE S.A.

**Table 2.3:** Market Volume, i.e. Energy and peak power demand (2005-2011) for the interconnected system (Source: IPTO - ADMIE)

|  | 2005 | 2006 | 2007 | 2008 | 2009 | 2010 | 2011 |
|---|---|---|---|---|---|---|---|
| Electricity consumption excluding pump storage (GWh) | 52,500.8 | 53,656.8 | 55,253.4 | 55,675.3 | 52,436.5 | 52,365.8 | 51,872.3 |
| Peak load (MW) | 9,635 | 9,962 | 10,610 (11,110 including curtailed load) | 10,393 | 9,828 | 9,902 | 10,055 |

**Figure 2.1** provides information on the main structural components of the Greek wholesale Electricity Market (taken from RAE's 2010 National Report to the European Commission) (RAE, 2009 and 2010).

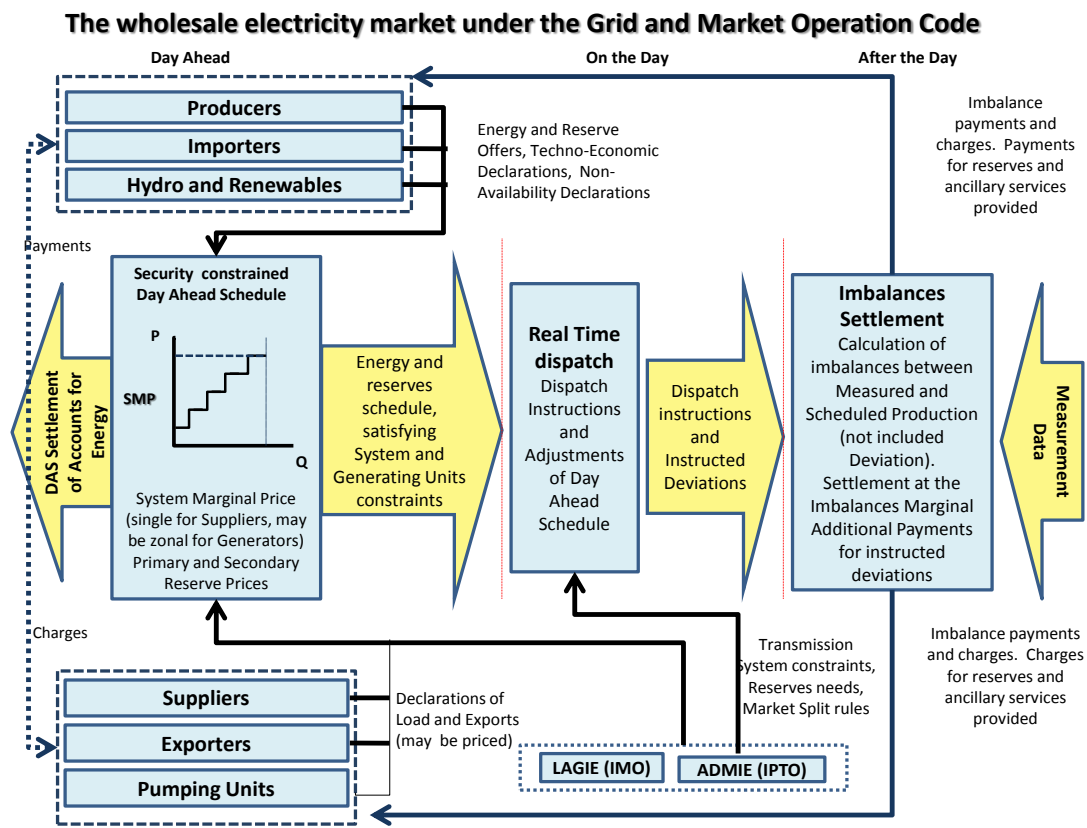

**Figure 2.1.:** Structural components of the Greek wholesale Electricity Market (taken from RAE's 2010 National Report to the European Commission)[4].

---

[4] *Amended by the authors to depict the formation of two new entities i.e. LAGIE and ADMIE, out of the former HTSO*

**Table 2.4: Hydroelectric Stations and River systems in GEM**

| Hydro Electric Station Dam | River | Total Installed Capacity (MW) |
|---|---|---|
| Kremasta | | 437 |
| Kastraki | Aheloos | 320 |
| Stratos | | 150 |
| Polyfito | | 375 |
| Sfikia | Aliakmon | 315 |
| Asomata | | 108 |
| Aoos Springs | | 210 |
| Pournari I | Araxthos | 300 |
| Pournari II | | 34 |
| Thisavros | Nestos | 384 |
| Platanovrisi | | 116 |
| Ladonas | Ladon | 70 |
| Plastiras | Tavropos | 130 |

2.2 **Regulatory Market Reforms RMRs**

Because one of the main targets of this work is the detection of the impacts of the various reforms made, until today, by the Regulator on the SMP, we consider necessary a small description of these reforms as this will help a lot in understanding the usefulness of the modeling results of this work. The reforms took place on specific dates -milestones or ***Reference Dates,*** within the aforementioned transitional phases. Regulator's National reports to EU (RAE, 2009,2010,2011) as well as other texts of the ADMIE and the work of Kalantzis et.al. mentioned above, are the main sources of this brief description.

However, in all the previously referred official reports of the regulator regarding the prevailing market conditions, it is a spread evidence that GEM exhibits significant structural distortions and weaknesses that influence negatively its efficiency. These frictions are responsible for the lack of real competitiveness in the market. As an example, Regulator has repeatedly pointed out in reports the need for measures to constraint the excessive market power of the dominant player in the market, PPC. The regulator considers that PPC behaves completely differently than the other rivals due to the fact that it has a unique energy mixture,(consisting of Lignite and Hydropower Stations that the other do not allow to have. This according to the regulator and other players creates an asymmetry in the market and subsequently frictions impairing efficiency.

More specifically, some of the measures taken by RAE towards avoiding abuses by PPC, are the transparency of information, Techno-economic declarations of generation units, periodical Hydro Usage declarations, Unit Availability declarations, Price Caps and Price floors (RAE, 2010). Given the fact that in general and in particular in GEM Hydro Generation has a crucial impact on the formation of SMP, RAE requested PPC to publish 12-month rolling estimates of mandatory hydro, reservoir inflows and storage levels and historical annual curves. Through this measure, RAE's purpose is to make possible deviations, between target and actual levels in inflows and storage, more clear and amenable to potential corrections. A measure that also gives to the traders the ability to adjust their positions in response to the published hydro schedules is the obligation of PPC to publish those schedules prior to the closure of the auctions for interconnection.

**1st Reference Day (1.10.2005) (RMR1).** The **1st Reference Day (01.10.2005)** marked the beginning of a market-based operation of the system, with the System Operator dispatching units according to an offer based unit commitment and the SMP calculation to be based on offers instead of costs.A price floor was set (it is still in place) to the offers, corresponding to each Unit´s minimum variable cost, in an effort to prevent the incumbent (dominant player generating) from bidding below this value. This reform in combination with the entrance in the market of a new IPP (ENTHES), was expected to have a positive effect (increase) on the evolution of SMP. To facilitate the dispatching of Units, based on at least their technical minimum, the price floor was not applied to the first 30% of the offered quality (rule of 30%).

**2nd Reference Day (1.1.2006) (RMR2). The Capacity Adequacy Mechanism, CAM**, aiming to the partial recovery of capital costs, forces Suppliers to buy capacity certificates from generators. In November 2010, until now, the value was set to 45000€/MWh/year, in an effort to mediate the effect of low demand on the revenues of generators. CAM has created incentives for new investments, inducing as almost 2000 MW of new IPP gas-fired capacity to be added to the system by the end of 2011.**CAM**, has an effect on SMP depending on the degree of maturity of the market (Stoft S., 2002) If the market is mature this mechanism is expected to have no effect on the dynamics of SMP. **The third Regulatory Market Reform, RMR3**, was a result caused by an almost continuous pressure by market agents to change the methodology of estimating SMP. In addition to this, there were also expressed complaints by market players for the operations of the System Operator regarding the dispatching of a number of Units, at all times, for security reasons, some of them functioning at their technical minimum. A new methodology was adopted aiming in changing the artificially low SMP determined due to the above mentioned reasons, causing the agents to complaint, in which technical minimum were not considered, therefore approaching to a pure economic dispatch. An increase in SMP was expected due to this reform. **RMR4** regards the change from daily to hourly submission of generator offers, starting on 4th April 2007. According to this reform each generator could submit for each trading day and for each hour, a different bid of higher value for peak hours and low for off-peak ones. The SMP was expected not to be affected by this reform.

**3rd Reference Day (1.5.2008) (RMR5). Cost Recovery Mechanism, CRM[5]**, was considered by the Regulator a necessary step until the **Imbalance Settlement Mechanism, ISM** (scheduled for the 5th Reference Day). CRM states that if the SMP is lower than the marginal cost of generating Unit (plus 10%), then the Unit will receive the difference as a compensation. The Regulator expected that this Reform would have no effect on SMP. CRM was aiming to ensure that generators will be compensated at least their marginal cost, in case they were ordered to operate.

**4th Reference Day (1.1.2009) (RMR6).** The **4th Reference Day (01.01.2009)** focused on the change of the ex-post SMP calculation methodology according to the unit commitment algorithm that considers all technical constraints of the units and the reserve requirements of the ADMIE (ex HTSO) expecting to lead to lower SMPs.

**5th Reference Day (30.9.2010) (RMR7).** The **5th Reference Day (30.09.2010)** initiated the mandatory day-ahead market model and introduced the Imbalances Settlement Mechanism retaining at the same time the SMP methodology allowing only the submission of demand declarations. **RMR7** is referred to the adoption of an enhanced Unit commitment algorithm which co-optimize energy as well as ancillary services. In this new mandatory, Day-Ahead market model incorporating, at the same time, an Imbalance settlement mechanism[6], market clearance is now based on the non- priced demand declarations (previously the HTSO forecasts were used instead). Taking into account that the methodology for estimating SMP retained the same and the fact that usually the declared demands were underestimated, the effect of this reform expected to reduce SMP slightly. **RMR8** regards the decision of the Ministry of Finance (1.9.2011) to impose a new tax levy on natural gas, equal to 1.50€/GJ (applied also to electricity generation). As SMP was set, for the majority of trading periods, by Natural Gas fired Units, the resulted increased generation cost was expected to increase SMP (see section 6.1 for comments). At the final stage the "architecture" of the market consists of the following components: **A Day-Ahead (DA) compulsory market, the Imbalances Settlement Mechanism,** the **Capacity Adequacy Mechanism,** the **Cost Recovery Mechanism** and the **Ancillary Services Market.** GEM is supervised by the Regulatory Authority for Energy (RAE) and is operated by the newly formed organizations, the Independent Power Transmission Operator (IPTO or ADMIE) and the Independent Market Operator (IMO or LAGIE). We provide a small description of these two bodies and a history of their formation process in the appendix 2. Table 2.5 summarizes all the Regulatory Market Reforms taken towards a resolution of the distortions. In this table we provide the values of SMP and Load at the time of enactment of each reform, in order to monitor the evolution of SMP returns and their volatility, after the enactment. The last two columns give RAE's expectations of the impacts these reforms may have on SMP and the actual, observed in practice impacts that also happened to be reassured by the modeling results of this paper. As we will see in a later section, for example, the Cost Recovery mechanism had actually a positive impact on the evolution of SMP, as it is shown in the last colum of Table 2.3, instead of an expected by the regulator neutral or uncertain one.

---

[5] This mechanism, currently in use, provides an explicit compensation for the commitment costs incurred as a result of the market situation (generation scheduling) as well as additional payments so the Unit ends up with a profit equal to 10% of its variable cost, if the market revenues for energy (taking into account both DA and IS markets) is not capable of generating a reasonable profit (see also Andrianesis P., et.al , 2011).

---

[6] All imbalances – referring to the differences between the DAS (Day-Ahead-Schedule) and the real production or withdrawal of electricity – are settled through the Imbalance Settlement Mechanism.

**Table 2.5: History of Regulatory Market Reforms (RMR)**

| Regulatory Market Reform (RMR) | Launching Date of RMR | Description of the Regulatory Market Reform | Values at time of Regulatory change | | Regulator's expected Impact of RMR on SMP) | Actual impact of the Reform |
|---|---|---|---|---|---|---|
| | | | SMP €/MWh | Load MW | | |
| RMR1 | 01.10.2005 | Adoption of mandatory pool (ex-post settlement) | 63.12 | 5322 | upward | upward |
| RMR2 | 01.01.2006 | Capacity Adequacy Mechanism | 27.46 | 4975 | neutral | neutral |
| RMR3 | 13.01.2006 | 1st Change of methodology for SMP calculation | 41.04 | 6839 | upward | upward |
| RMR4 | 01.04.2007 | Introduction of Hourly Bids | 41.98 | 5292 | neutral | downward |
| RMR5 | 01.05.2008 | Cost Recovery Mechanism | 42.02 | 4544 | neutral | upward |
| RMR6 | 01.01.2009 | 2nd Change of Methodology for SMP calculation | 47.97 | 5538 | downward | downward |
| RMR7 | 30.09.2010 | DA market (mandatory) with imbalance settlement | 52.79 | 5532 | downward | upward |
| RMR8 | 01.09.2011 | Introduction of Natural Gas Consumption Tax | 90.75 | 6509 | upward | upward |

### 3. Data Sets, Processing and Tests

We have selected and analyzed daily data (average of 24 hourly data) on spot, Ex-post System - wide Marginal prices, SMP, and load (demand) of the Greek Electricity Market, taken from the official site of the Greek Independent Power Transmission Owner, IPTO, ADMIE S.A (www.admie.gr). The data set cover the period from 01.01.2004 to 31.12.2011, i.e. it contains 2922 data corresponding to eight years. The market was less mature further back. Due to transformations needed for some variable (e.g Brent Oil data), the final length of the data set used in all calculations is 2877. We note that the SMP data for the period 01.01.2004 to 30.09.2010 are **ex-post** adjusted values while from 01.10.2010 to 31.12.2011 are **ex-ante**. Figure 3.1 shows the evolution dynamics of SMP values. An upward trend which drives the series is evident, causing an increase of X% in the period 2004-11. Regular (weekly, quarterly) patterns can be seen, as well as some short-lived spikes and **volatility clusters** (although not easily distinguished). We clarify, at this point, the characteristics of the chosen series for analysis and modeling.

During the transitory market regime (2004 – Sept. 2010), an indicative plant – commitment schedule was formed by the Day – Ahead market which also determine a SMP forecast as a reference spot price, working mainly as a price signal. However, the Cash – flows were based on **Ex–post SMP** prices which were the output of solving again the same cost – minimization problem as in the case of Day – Ahead Schedule (DAS). This was possible by considering, in the solution process, the actual metered values of inputs as demand (load), plant availabilities, RES generation etc, and not DA forecasts. Then, Ex–post SMP prices were applied to the actual generation (reflecting largely the real – time dispatch orders of the TSO) and Consumption Volumes. Therefore, in the transitional period of GEM, Ex–post SMP used to settle the whole market. After the RMR7 (30.09.2010), the new market design consists of two separate settlement processes: the **DA market settlement** (in which generators are paid and suppliers are charged according to volumes based on SMP and the plant schedules formed from DA dispatch i.e. from load declarations submitted, and the **Settlement of Imbalances.** In this settlement, deviations from DAS are charged or compensated (exogenous or TSO dispatch orders). In DA market a uniform pricing is still in place, which reflects the offer of the most expensive Unit dispatched, taking into Consideration the forecasted demand, plant technical constraints and the needs for reserve capacity. Table 2.1a gives the summary statistics of the two prices . Figures 3.1a ,b show the co-evolution dynamics and the distribution comparison.
In the box plot, the values of the SMP ex-post distribution parameters (3$^{rd}$ quartile, mean, median and 1$^{st}$ quartile) are slightly higher than those of SMP-das. However the number of extreme or outlier values is larger and more extensively distributed in the case of SMP ex-post.

Table 3.1a: Summary statistics of SMP_das and SMP_expost

|  | DAS | EXPOST |
|---|---|---|
| Mean | 56.35368 | 61.55577 |
| Median | 54.80146 | 59.81215 |
| Maximum | 117.9084 | 150.0000 |
| Minimum | 20.57900 | 20.37208 |
| Std. Dev. | 18.05566 | 19.24139 |
| Skewness | 0.442226 | 0.655339 |
| Kurtosis | 2.804380 | 3.824370 |
|  |  |  |
| Jarque-Bera | 87.38570 | 255.3301 |
| Probability | 0.000000 | 0.000000 |
|  |  |  |
| Sum | 144040.0 | 157336.5 |
| Sum Sq. Dev. | 832947.1 | 945940.1 |
|  |  |  |
| Observations | 2556 | 2556 |

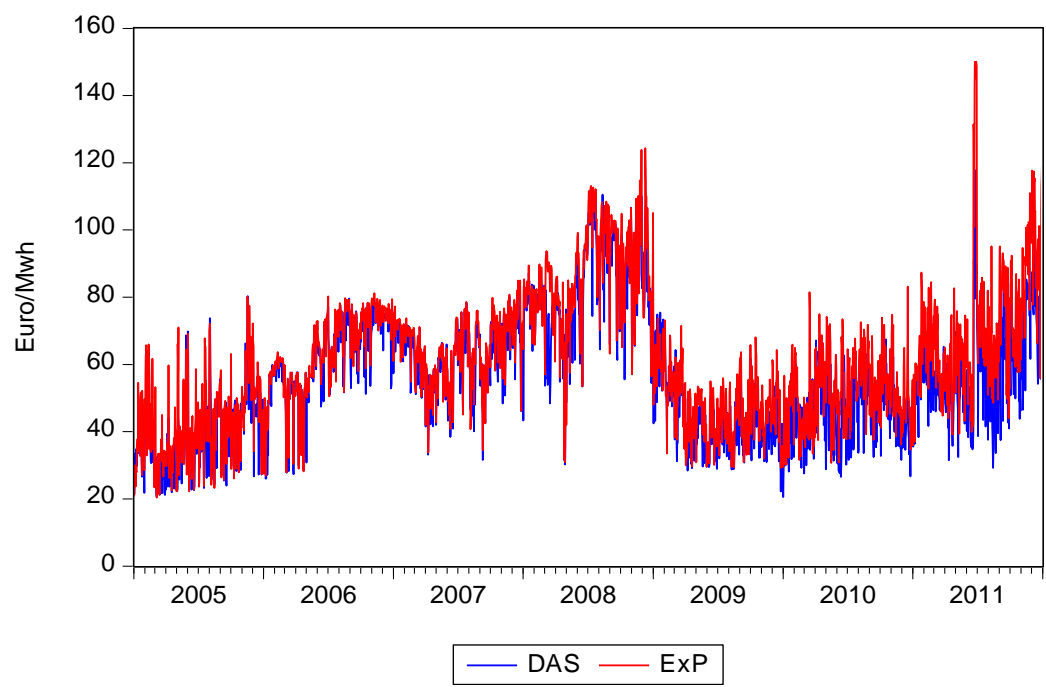

Figure 3.1a Time histories of SMP Das and SMP ex-post, 2004-11.

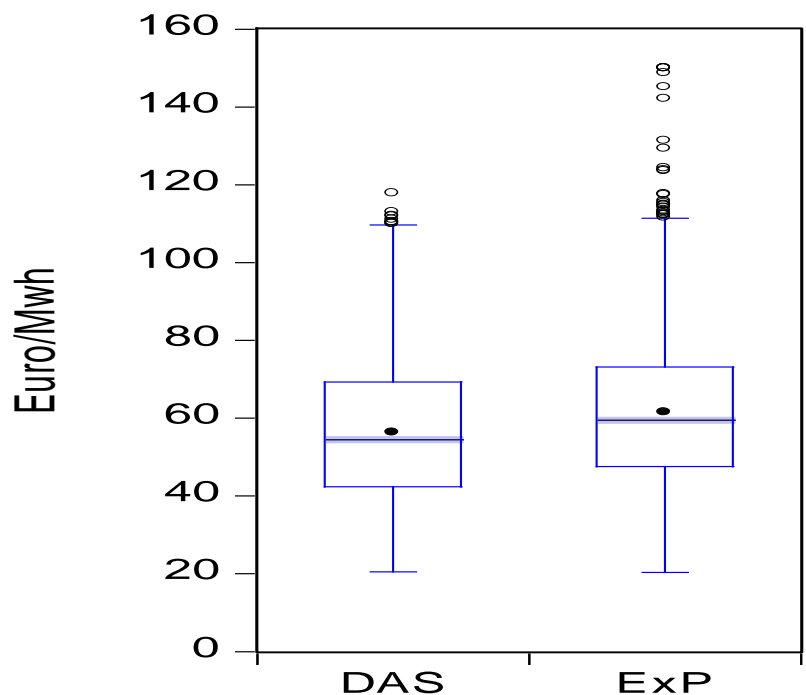

Fig. 3.1b Box plot of SMP Das and Ex-post .

As temperature is probably the most commonly used load predictor which in turn influences heavily SMP we examine the (linear) correlation coefficient between load and minimum, maximum temperature data from meteorological stations installed by NTUA, National Technical University of Athens. Table 2.1b shows daily max  and  min temperatures in Thessaloniki (the second largest city of Greece, in North) and  in Athens  (South). We observe that until 2009, North and South , load is correlated with min and max temperature almost the same. However, during the last two years 2010-2011, the correlation becomes more intense, both in two regions, while in 2011 in Athens we observe a greater correlation of load with min temperature indicating a more intense usage of AC for heating during winter due to a more expensive price of heating Oil and Nat Gas (a tax was imposed by the Government). The variation of Load with the daily average temperature in the country is shown in **figure 3.2.**

**Table 3.1b** Correlation coefficient between daily average Load and Temperature

| Time Period | Correlation coefficient between daily average Load and Temperature | | | |
| --- | --- | --- | --- | --- |
| | THESSALONIKI | | ATHENS | |
| | min temp. °C | max temp. °C | min temp. °C | max temp. °C |
| Full series | 0.29 | 0.29 | 0.35 | 0.32 |
| 2007 | 0.16 | 0.15 | 0.17 | 0.16 |

| 2008 | 0.14 | 0.11 | 0.21 | 0.18 |
| 2009 | 0.14 | 0.17 | 0.19 | 0.19 |
| 2010 | 0.43 | 0.44 | 0.48 | 0.48 |
| 2011 | 0.42 | 0.43 | 0.48 | 0.44 |

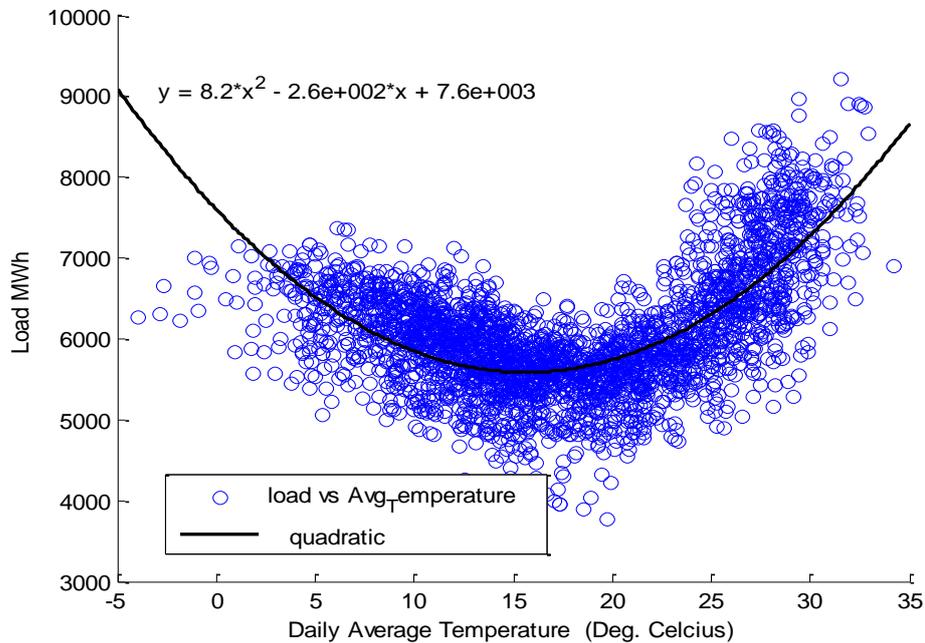

**Figure 3.2** Load variation with average daily temperature in Greece. In the quadratic equation shown x is temperature and y Load.

The graph shows a parabolic shape indicating increased consumption at high and low temperatures. It suggests also that forecasting future load (demand) requires the knowledge of load growth profile, relative to a certain reference i.e the current year. The polynomial function of load versus temperature shown on the graph seems a reasonable approximation for load forecasting. Due to quadratic and strong correlation between temperature and load, we have not included temperature as an exogenous – regressing variable in the modeling since its effect on SMP is sufficiently captured by the load variable.

The data for **Imports** and **Exports** of Energy (in MWh) were taken from interconnection measurements (per interconnection point). The Generation Thermal Units **availability** data are daily final while for the availability of Hydro Units we considered their contractual values.
The management of some missing values in the SMP spot price vector was done by using a combination of approaches like average of the neighboring observation for sporadic missing values. An overview of the data is given in Table 3.2.

**Table 3.2:** List of data set and name of Variable used in Modeling.

| | Name of Data TS | Description | Length of Time Series | Units of Measure | Resolution Daily | Resolution Monthly | Source | Period Covered |
|---|---|---|---|---|---|---|---|---|
| 1 | SMP ex-post or SMP | Ex-post System Marginal Price (GEM Pool) | 2877 | €/MWh | ✓ | ✓ | IPTO Data Base | 2004-2011 |
| 2 | Load | Load or Demand | 2877 | MW | ✓ | | IPTO Data Base | 2004-2011 |
| 3 | Brent | Brent Crude OIL Price | 2877 | $/pbl | ✓ See note | | ICE (Inter. Exchange, Blooberg 2004-2012) | 2004-2011 |
| 4 | HydMR | Hydro Production, must-run | 2877 | MWh | ✓ | ✓ | IPTO Data Base | 2004-2011 |
| 5 | HydGen | Hydro Production | 2877 | MWh | ✓ | ✓ | IPTO Data Base | 2004-2011 |
| 6 | Ahelenr | Energy Reserves of Aheloos Dam | 2877 | MWh | ✓ | ✓ | IPTO Data Base | 2004-2011 |
| 7 | Aliakenr | Energy Reserves of Aliakmon Dam | 2877 | MWh | ✓ | | IPTO Data Base | 2004-2011 |
| 8 | Araxthenr | Energy Reserves of Arahthos Dam | 2877 | MWh | ✓ | | IPTO Data Base | 2004-2011 |
| 8 | Ladonenr | Energy Reserves of Ladon Dam | 2877 | MWh | ✓ | | IPTO Data Base | 2004-2011 |
| 10 | Nestosenr | Energy Reserves of Nestos Dam | 2877 | MWh | ✓ | | IPTO Data Base | 2004-2011 |
| 11 | Plastenr | Energy Reserves of Plastiras Dam | 2877 | MWh | ✓ | | IPTO Data Base | 2004-2011 |
| 12 | Hfog | Heavy Fuel Generation | 2877 | MWh | ✓ | | IPTO Data Base | 2004-2011 |
| 13 | Resgen | Renewables Generation | 2877 | MWh | ✓ | | IPTO Data Base | 2004-2011 |
| 14 | Ligng | Lignite Generation | 2877 | MWh | ✓ | | IPTO Data Base | 2004-2011 |
| 15 | Natgasg | Natural Gas Generation | 2877 | MWh | ✓ | | IPTO Data Base | 2004-2011 |
| 16 | TotImports | Total Imports of Energy | 2877 | MWh | ✓ | ✓ | IPTO Data Base | 2004-2011 |
| 17 | TotExports | Total Exports of Energy | 2877 | MWh | ✓ | ✓ | IPTO Data Base | 2004-2011 |
| 18 | Uvail | Generating Unit Availability | 2877 | MWh | ✓ | | IPTO Data Base | 2004-2011 |
| 19 | MinTempNTUA MaxTempNTUA AvgTempNTUA | Min, Max, MeanTemperature in Athens | 2877 | °C | ✓ | ✓ | National Technical University of Athens, NTUA | 2004-2011 |

Figure 3.3 and 3.4 show the SMP (raw data) and load (demand) time histories. It is apparent in fig. 3.3 that SMP exhibit a characteristic known as **volatility clustering**, that large changes and small changes tend to have their own clusters. The concept of volatility clustering is further explained in the sections that follow.

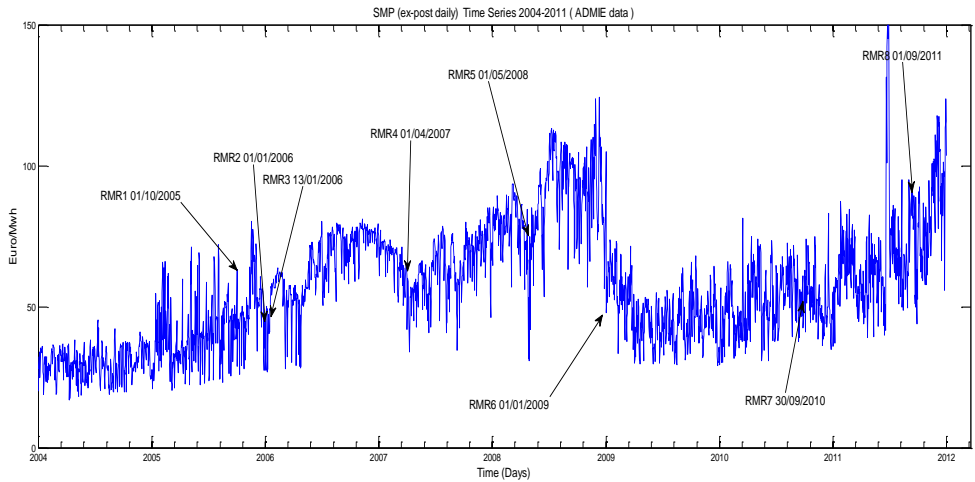

**Figure 3.3.:** Time history of the ex-post System Marginal price, SMP, of the Greek Electricity Market, 2004-2011. The arrows locate the dates of regulatory changes.

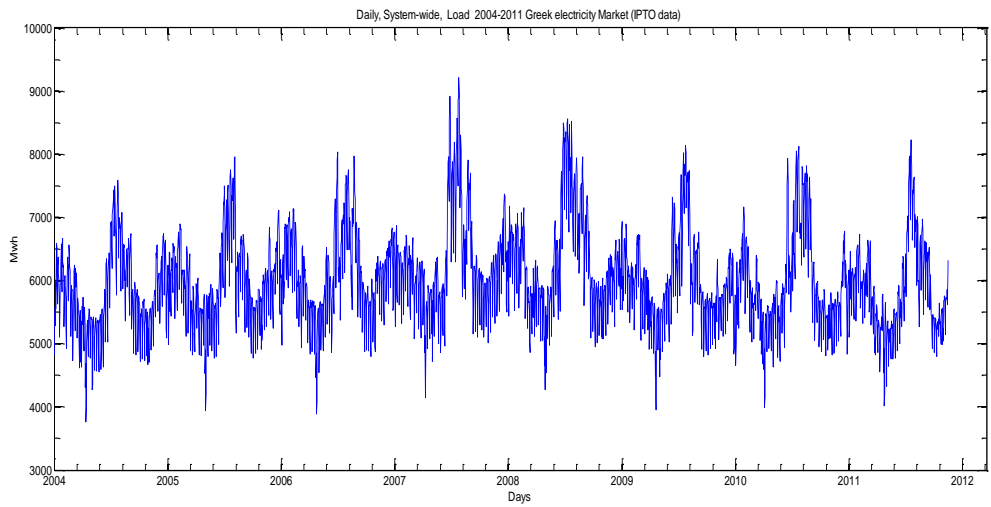

**Figure 3.4.:** Time history of the daily, system-wide, load of the Greek electricity market 2004-2011.

**Table 3.3:** Summary statistics of SMP

| Period | Min Euro/Mwh | Max Euro/Mwh | Mean Euro/Mwh | Std Deviation Euro/Mwh | Coefficient of Variation (a Volatility proxy) | Skewness | Kurtosis | Inter Quartile Range Euro/Mwh |
|---|---|---|---|---|---|---|---|---|
| 2004 | 16.97 | 45.19 | 30.00 | 5.22 | 0.17 | -0.1578 | 2.614 | 7.22 |
| 2005 | 20.37 | 80.16 | 43.13 | 11.97 | 0.28 | 0.5246 | 2.986 | 15.66 |
| 2006 | 26.91 | 81.19 | 64.13 | 12.06 | 0.19 | -0.8428 | 3.5010 | 18.91 |
| 2007 | 33.82 | 84.97 | 64.94 | 9.27 | 0.14 | -0.6681 | 3.3435 | 12.93 |
| 2008 | 30.92 | 124.35 | 87.23 | 14.23 | 0.16 | -0.4520 | 3.999 | 19.12 |
| 2009 | 29.19 | 73.66 | 47.39 | 9.67 | 0.20 | 0.3630 | 2.7483 | 12.91 |
| 2010 | 29.42 | 83.12 | 52.25 | 10.51 | 0.20 | 0.1783 | 2.4992 | 15.07 |
| 2011 | 35.94 | 150 | 71.70 | 20.11 | 0.28 | 1.4048 | 5.800 | 23.07 |
| 2004-2011 | 16.97 | 150 | 57.00 | 20.36 | 0.36 | 0.540 | 3.464 | 30.25 |

Table 3.3 gives the descriptive statistics of daily SMP. The positive skewness and kurtosis of the entire series indicate that the distribution of SMP has an asymmetric distribution with a longer tail on the right side, fat right tail. SMP is clearly non-normally distributed, as it is indicated by its skewness and excess kurtosis (>3) (normal distribution has kurtosis ≈3).

Table 3.4 provides summary statistics for SMP and log SMP (the choice of this transformation is explained below) together with results from normality tests applied on the series, using the **Jarque – Bera, JB** algorithm (Jarque C.M. and A.K. Bera, 1987) of the null hypothesis that the sample SMP has a normal distribution. The test rejected the null hypothesis at the 1% significance levels, as it is shown in the table (p=0.000, JB Stat significant). As it is shown in a later section, SMP follows a distribution of fat tail, a student's t distribution.

Table 2.4 Comparing summary statistics of SMP and log SMP

|  | SMP | LOG(SMP) |
|---|---|---|
| Mean | 56.98370 | 3.975591 |
| Median | 56.07000 | 4.026601 |
| Maximum | 150.0000 | 5.010635 |
| Minimum | 16.97000 | 2.831447 |
| Std. Dev. | 20.36816 | 0.375986 |
| Skewness | 0.540037 | -0.371518 |
| Kurtosis | 3.464248 | 2.650906 |
|  |  |  |
| Jarque-Bera | 165.6774 | 80.79185 |
| Probability | 0.000000 | 0.000000 |
|  |  |  |
| Sum | 163942.1 | 11437.77 |
| Sum Sq. Dev. | 1193142. | 406.5675 |
|  |  |  |
| Observations | 2877 | 2877 |

In figures 3.3 the SMP exhibits also, besides volatility clustering, the typical features of **mean reversion** and **spikes,** a tendency of the data to fluctuate around a long-term stable state or equilibrium, as well as extremely high values of short duration (spikes). Figure 3.5 shows the quantile-quantile plot of SMP against a theoretical normal distribution, In case the empirical and assumed theoretical distributions are the same, the q-q plot should be a straight line. The deviation of SMP from normality is apparent in the q-q plot. We see that the empirical quantiles (blue curve) versus the quantiles of the normal distribution (red line) do not coincide, even slightly, and exhibit extreme **right and left tails**: the maximum for SMP data is a multiple of 4.8 of the standard deviation, often far larger than the factor around 3.46 that the normal distribution suggests. The extreme right tails of SMP will be used later as a way of describing the **spikes** that are shown in the conditional volatility series. Summary statistics of all the exogenous quantitative explanatory variables, used in modeling process is given in the Appendix 3 of this paper.

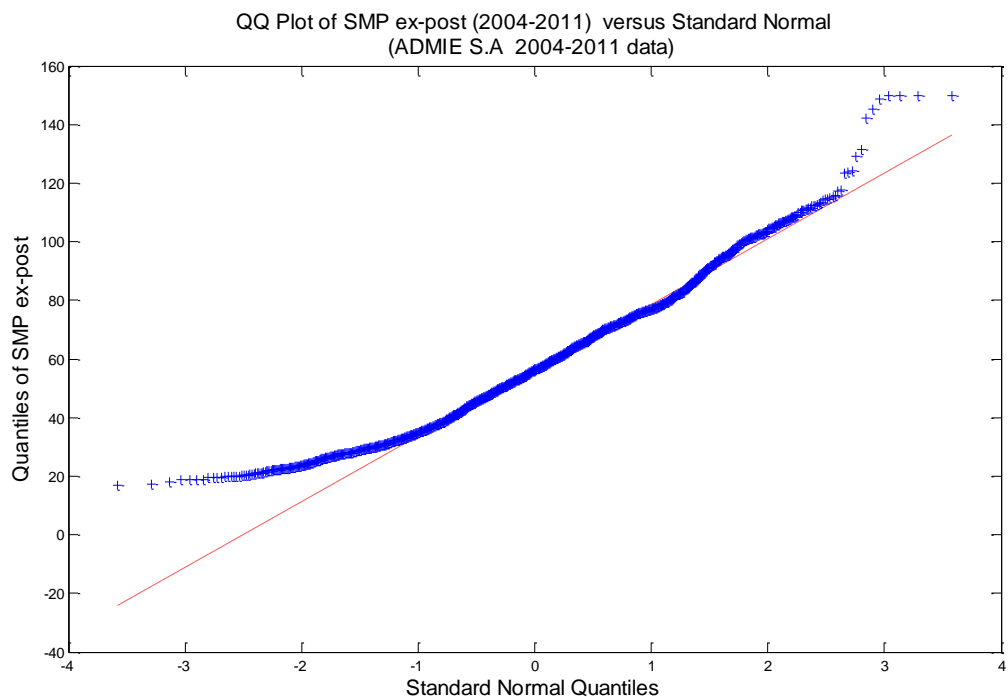

**Fig. 3.5:** Quantile – Quantile plot of ex-post SMP against the normal distribution

### 3.3 Testing for stationarity of the SMP returns

We treat SMP as a stochastic process and before proceeding to further in our analysis, it is necessary to check the series for lack of stationary. Economic and Financial time series, due to the fact that they depend on exogenous factors, exhibit a non-stationary behavior (Brock A.W. and P.J.F. de Lima, 1995, Pagan A., 1996). However, it has been found that electricity market data are more stationary than all financial series, reported so far in empirical analyses (Strozzi F. et. al., 2002, Bunn 2004, Weron R., 2006). Augmented Dickey_Fuller (ADF) and Phillips-Perron (PP) unit roots tests have been performed, for testing the null hypothesis that SMP has unit roots i.e is nonstationary. Table 2.5 lists the results.

Table 3.5 ADF and PP test for Normality.

| Lag | h | p | ADF Stat | h | p | PP test |
|---|---|---|---|---|---|---|
| 0 | 1 | 0.000 | -12.9068 | 1 | 0.000 | -12.9068 |
| 1 | 1 | 0.000 | -10.4557 | 1 | 0.000 | -11.9549 |
| 2 | 1 | 0.000 | -8.6783 | 1 | 0.000 | -11.2853 |
| 3 | 1 | 0.000 | -7.6094 | 1 | 0.000 | -10.9366 |
| 4 | 1 | 0.000 | -6.8322 | 1 | 0.000 | -10.7832 |
| 5 | 1 | 0.000 | -5.9425 | 1 | 0.000 | -10.6480 |
| 6 | 1 | 0.000 | -5.0524 | 1 | 0.000 | -10.5460 |
| 7 | 1 | 0.000 | -5.6615 | 1 | 0.000 | -10.8849 |

The logical flag h = 1 indicates a rejection of the null hypothesis (critical Value for both tests is -3.414 a value less than the Stats values above). Therefore SMP with high probability is covariance-stationary around a linear trend. Theodorou (Theodorou P., et. al, 2008, Petrell et. al, 2012) have reported similar with us results using Dickey – Fuller (DF) and Kwiatkowsk – Phillips – Schmidt – Shin (KPSS) tests. Papaioannou G., et al. (1995) and Papaioannou G., (2013), have applied a nonlinear tool for stationarity detection in financial and electricity markets.

ADF unit root tests reject the null hypothesis of the existence of this type of nonstationarity for all Hydro Plants (energy reserves series) and the Brent oil one (ADF Stats > CriticalVal). Therefore, for these series we take first differences to make them stationary. SMP and all quantitative explanatory variables are not Normally distributed as the JB test suggests, so we log transform them to make them 'more normal' (JB Stats are reduced and become insignificant) (see Appendix 3 for test results).

### 3.4 Autocorrelation and Partial Autocorrelation functions

Autocorrelation estimates of SMP and its returns are presented in this section. They signal the extent to which prices carry over from one day to the next. Arbitrage opportunities are associated with non-zero autocorrelations.

In figure 3.6 we plot the ACF of the **system marginal** price for the period 2004-2011. We observe **strong, persistence 7-day dependence and that** the ACF decays very slowly. This indicates that SMP data contain a strong trend which is responsible for this positive and very slowly decreasing ACF. Brownian motion, commodity prices etc. behave similarly, yielding very slowly decaying ACF's. However, these serial dependencies can be "destroyed" by taking differences in the data which is equivalent of taking the returns of the data. Since we are interested in revealing the impacts of exogenous variables (that exhibit both trend and periodicity) on SMP and since SMP series was found to be stationary we do not take differences of the series and let our proposed composite ARMAX/GARCH model to take care of these dependencies.

We have also calculated ACF and PACF (not shown here) of SMP returns and square returns (a proxy for variance) in which we see also that there is a strong, persistent of one-, two-, seven- and 14-day dependency in contrast to most financial data for which ACF of returns dies out quickly to zero after the first lag or after 5-10 lags (e.g. days) and long-term autocorrelations are observed only for squared returns (results are not included here due to space limitation). While the existence of such large autocorrelations indicates lack of market efficiency, the inherent characteristic of non-storability of electricity does not allow traders to take advantage for profits from such day-to-day dependence. The one- and two-day autocorrelations are due to the timing of bids. For example, bids for tomorrow delivery must be submitted today. So, the information set available at the time tomorrow's bids are submitted, includes yesterday's information but not today's one, therefore a two-day lag autocorrelation. Also, electricity demand varies from Sunday to Monday, and this variation is almost the same. The expected day-of-the-week effects observed in all electricity markets explain the 7- and 14-day lags in the ACF.

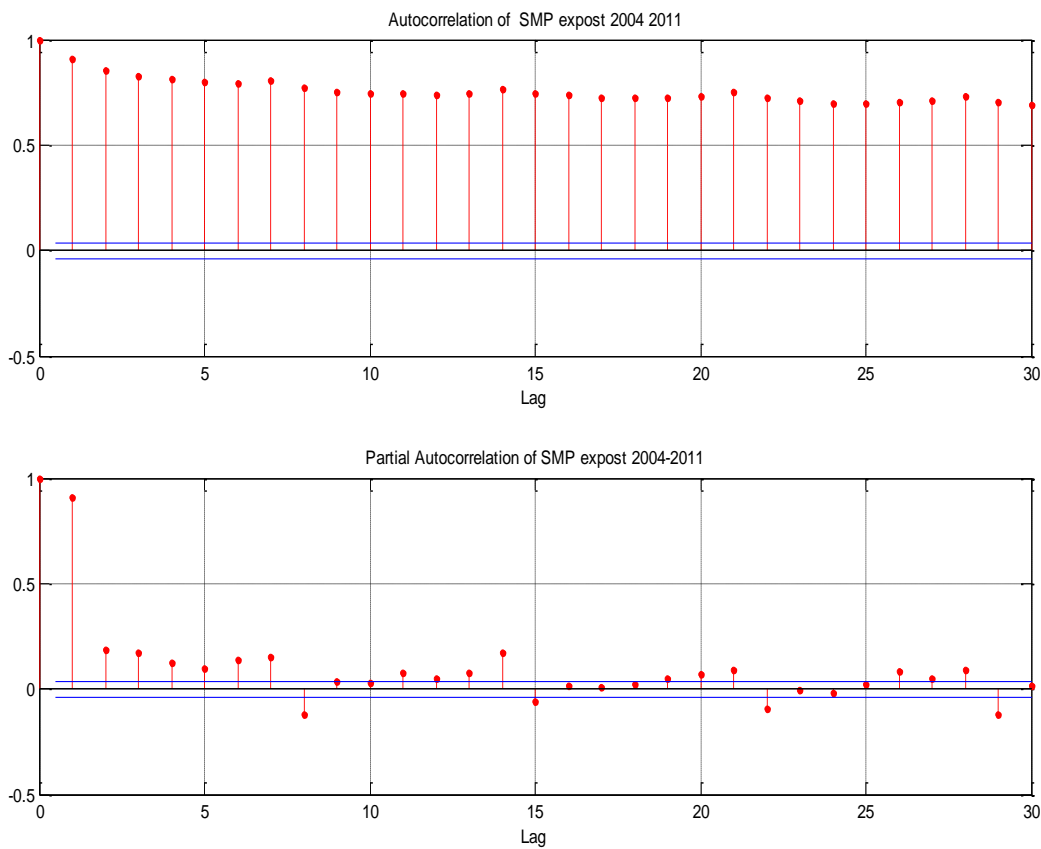

**Fig. 3.6 : Autocorrelation and** Partial Autocorrelation function of SMP (2004-2011)

The results obtained in figure 3.6 are useful in model specification for the **conditional mean.** A quantification of the correlation existence is possible by using a formal hypothesis test, the Ljung-Box-Pierce Q-test for a departure from randomness based on the ACF and PACF on SMP. It is known that the Q-test is most often used as a post-estimation lack-of-fit test applied to the fitted innovations (residuals)(see below). We have applied the above test using seven lags (weekly dependence) to test the null hypothesis of no serial correlation, under which the Q-test statistics is asymptotically $x^2$ distributed.

Table 3.6 Lung-Box-Pierce Q-test results on SMP

| lag | ACF | PACF | Q-Stat | prob |
| --- | --- | --- | --- | --- |
| 1 | 0.907 | 0.907 | 2366.9 | 0.000 |
| 2 | 0.854 | 0.183 | 4469.9 | 0.000 |
| 3 | 0.828 | 0.168 | 6443.5 | 0.000 |
| 4 | 0.809 | 0.118 | 8332.5 | 0.000 |
| 5 | 0.797 | 0.099 | 10162 | 0.000 |
| 6 | 0.795 | 0.134 | 11985 | 0.000 |
| 7 | 0.802 | 0.152 | 13843 | 0.000 |

Based on the results, the null of zero serial correlation is rejected at the 99% significance level.
We have also implemented Engle's test to test for the presence of ARCH effects. Under the null hypothesis that a time series is a random sequence of Gaussian disturbances (i.e. no ARCH effect exist), this test statistics is also asymptotically $x^2$ distributed. This test too rejected the null hypothesis that there are no significant GARCH effects (the results are not shown here). As a conclusion, the SMP series exhibit both significant correlations and **GARCH effect** or **heteroskedasticity.** This enhances our decision to apply a GARCH model for conditional volatility and its associated stylized effects.

Therefore, we are sure with a high certainty that the SMP series is covariance-stationary around a linear trend. Our result is in line with other previous works on electricity prices (Petrella, et. al,2012, Bunn, 2004, Werron, 2006). However, since we have found that SMP is not normally distributed we just take the log of the series to make it 'more normal' (actually the Jarue-Bera statistics see table 2.4 is reduced from 165.7 to 80.7).

Thus, based on these results, our models for conditional mean and volatility will be specified in log levels, that is we take log-transform and no differences. We expect that the linear trend mentioned before, may be nothing but a proxy for the underline trend in a fundamental factor as for example fuel prices. In fact, looking at the correlation table, between SMP and exogenous variables in Appendix 4 and the following figure 3.7 showing the Price Index changes in 2002-2009, one can expect that the linear trend may come from Nat Gas prices (correlation coefficient 0.57), Brent price (corr. Coef. 0.62) and load demand (corr. Coeff. 0.40). The **conditional mean** equation described below will give as a more certain answer for the linear trend, around which the SMP is covariance-stationary,

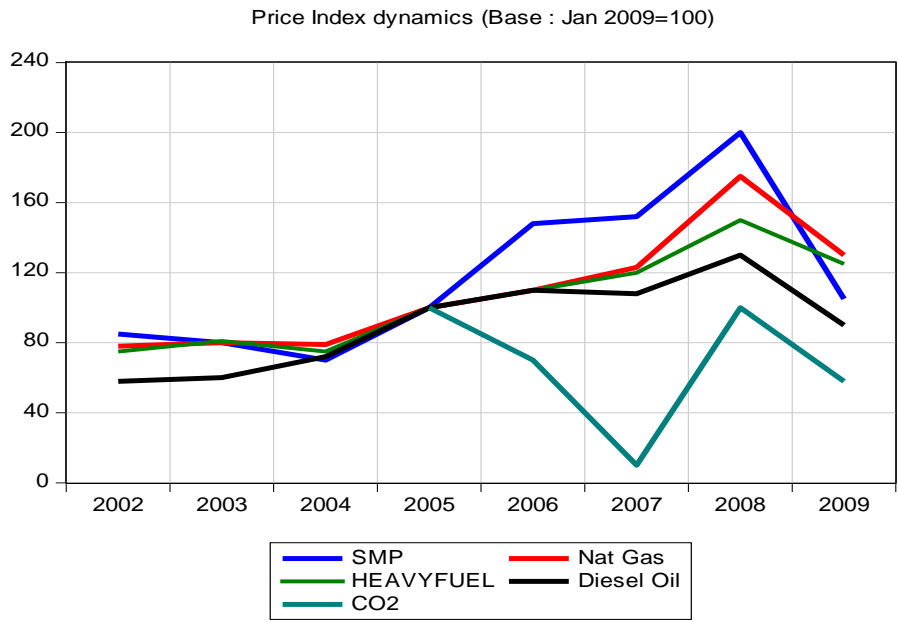

**Figure 3.7** Price Index dynamics of SMP and Fuels (source :FEIR report 2010[7])

---

[7] Foundation for Economic & industrial Research (IOBE), report 2010.

## 4 Modeling conditional mean and Unconditional (Historic) and Conditional Volatility of SMP

**Historical or statistical volatility, HV,** looks backward i.e. it is based on past return values, unlike **implied volatility** which is a concept used in the option pricing theory, giving an estimation for the future volatility. HV is simply the annualized **standard deviation, St.dev,** of past or historical returns. HV or retrospective volatility index therefore captures the amplitude of price movements for a given period of time in the past. All typical risk theoretical models that rely on the notion of standard deviation generally assume that return conform to a normal bell-shaped distribution. In that case, therefore, we can expect that about two-thirds of the time (68,3%), asset returns should fall within one st.dev (+-) while 95% of the time they should fall within two st.dev. Qualitatively normal distributions exhibit skinny "tails" and perfect symmetry. Using the **rolling volatility method** we estimate a rolling standard deviation i.e. the st.dev of returns measured over a subsample which moves forward through time, thus estimating volatility at each point in time. It is assumed implicitly in this approach that volatility is constant over the time interval corresponding to the chosen subsample. This method provides accurate values of st.dev, when we are interested in point estimation, and assumes also that returns follow a geometric Brownian motion (Merton, R., 1980).

However the assumption of constant volatility over some period is both statistically inefficient and inconsistent, in estimating volatility. We handle this problem by building first parametric models for the time-varying st.dev and then extract volatility from return data generated by the model. The models described in section 10 below (of the GARCH-type) are used to give better estimates of volatility of SMP.

The **moving window approach or rolling volatility approach** gives the volatility estimate at a given time $t_k$ (Weron R., 2006, Eydeland et al., 2003).
In figure 4.1 we show a 30-day and 90-day moving window or rolling volatility, for log SMP, for year 2011. A window of 30-day is also chosen in RAE's reports, and compare this unconditional volatility with the conditional one generated by the ARMAX/GARCH model described below.

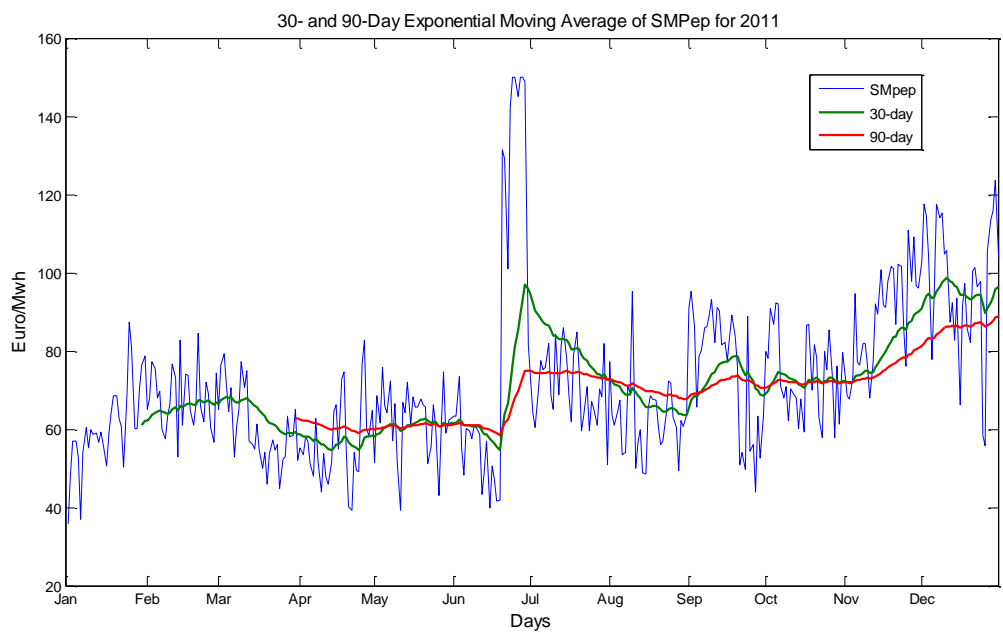

**Figure 4.1:** Dynamics of SMP ex-post for 2011(actual and 30-day and 90-day smoothed values by Exponential moving average of daily data) (Source: ADMIE S.A).

**Table 4.1:** 30-Day, m=30, Annualized Volatility for ex-post SMP returns

|  | 2009 | | 2010 | | 2011 | |
|---|---|---|---|---|---|---|
|  | **Volatility** | **Annualized Volatility *** | **Volatility** | **Annualized Volatility** | **Volatility** | **Annualized Volatility** |
| **Min** | 0.094 | 0.32 | 0.084 | 0.29 | 0.121 | 0.41 |
| **Max** | 0.220 | 0.76 | 0.280 | 0.97 | 0.291 | 1.01 |
| **Mean** | 0.154 | 0.53 | 0.180 | 0.62 | 0.184 | 0.64 |
| **std** | 0.026 | 0.09 | 0.037 | 0.13 | 0.039 | 0.13 |

*Annualized volatility = σ * $\sqrt{(365/m)}$ where we assume 365 trading days in GEM and m is the length of moving window.

However in the simple MA model, the moving averages are **equally weighted** (last day's return is equally weighted with the returns one week ago). In this "historic" model and all variations are due only to the differences in samples, i.e. a short moving average window e.g. m=10 days will be more variable than an m=30 one. But no matter which m is used, a simple rolling volatility model estimates the **unconditional** volatility which is just one number, a constant, underlying the entire series. In this model the wrongly perceived time variation in volatility is in reality a sample variation and there is no stochasticity estimation. No dynamic evolution of returns is taken into account by this model.

The equal weights on each return data are responsible for the appearance of "ghost features" in the volatility series. Therefore, instead of using a simple, rolling moving average model we use an **exponentially weighted moving average (EWMA)** that eliminates the "ghost features". A EWMA is written as follows:

$$\sigma_t^2 = (1 - \lambda)r_{t-1}^2 + \lambda \sigma_{t-1}^2 \qquad (4.1)$$

where λ is a smoothing constant. The larger the value of λ the more weight is placed on past data, the more smoother the series becomes.

Model (4.1) is used by JP Morgan-Reuter's Risk Metrics™ (J.P Morgan/Reuters, 1996) extensively in estimating volatility in financial assets (for daily data λ=0.94 is used). A careful look on (4.3) shows that (4.1) is actually a model for the **conditional volatility.** In fact, model (4.1) applied on squared returns is equivalent to a non-stationary GARCH model (see formula 4.7) IGARCH (Integrated GARCH), where k=0 and $A_1+G_1=1$. So, in order to compare the conditional volatility estimation by our best found ARMX/GARCH model, describe in section 6, with a "historic" EWMA model that also estimates conditional volatility, we construct an EWMA model which can produce as close as possible the behavior of our GARCH model. We use λ=0.94, as it is suggested by Risk Metrics™ (Risk Metrics™, 1996), in the technical document, equal to the **persistence coefficient $G_1$**, (the moving **window length** or **lagged time period** in EWMA is 77 days for 1% tolerance level). The dynamics of the generated conditional volatility series matches extremely close the GARCH's model dynamics, although differences in the levels is observed, as expected due to differences in model specification.

Large $G_1$ indicate that shocks to conditional variance take a **long time to die out**, so volatility is "persistent". Large $A_1$ mean that volatility is quick to react to market movements and volatilities show to be more "spiky". Even though there are differences in volatility values produced by the two models, their dynamic behavior is very similar (see fig. 4.7 for more details).

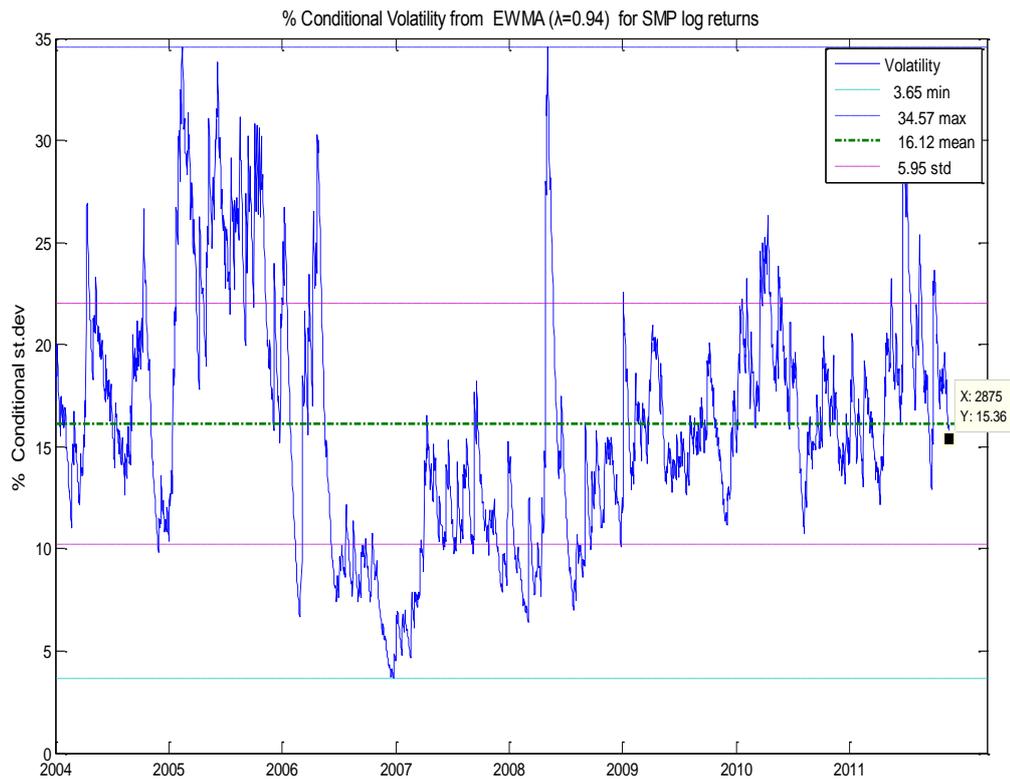

Figure 4.3 Conditional volatility estimated by EWMA model ($\lambda$=0.94)

Figure 4.4 shows the conditional correlation between SMP and load , the most significant driver of the volatility of SMP as the table of the coefficients of the fitted model shows. A very interesting point here is that the conditional co-movement of these two variables, reflected through their conditional correlation is gradually reduced after 2008, the year of the country's recession starts.

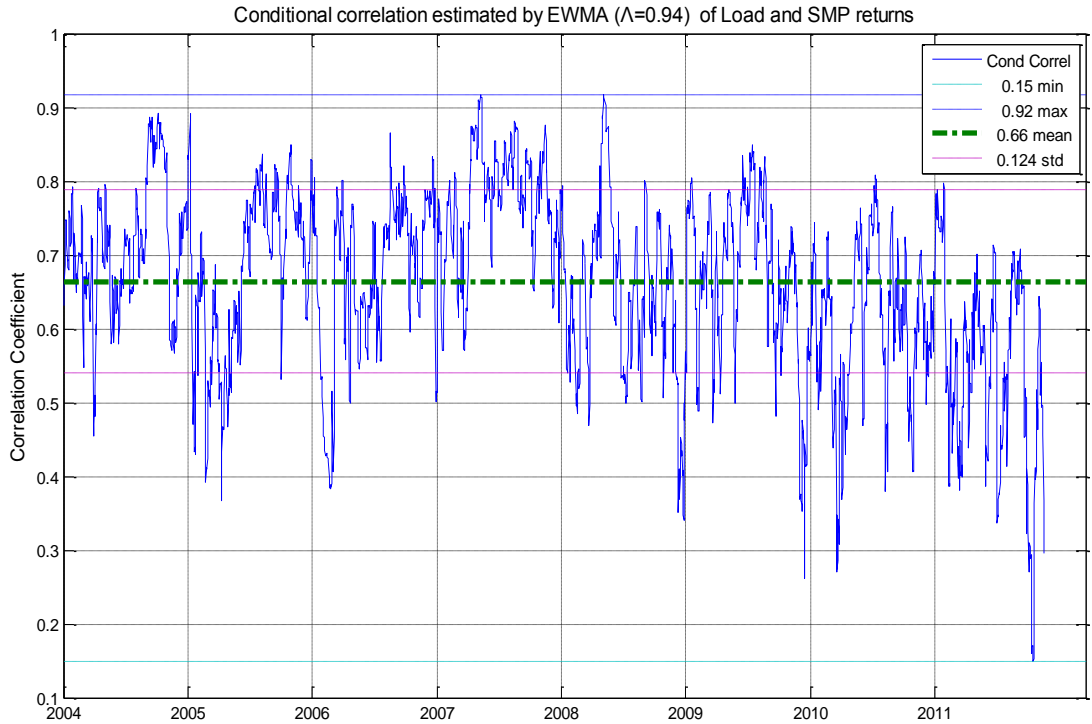

Figure 4.4 Conditional correlation between load (demand) and SMP estimated by EWMA model.

## 4.4 ARMAX/GARCH time Series Models for SMP

Volatility forecast models of the GARCH family are in general use. ARCH (Autoregressive Conditional Heteroskedasticity) of Engle (Engle R.F., 1982) was the first model but the generalized ARCH or GARCH (Bollerslev T., 1986) is the common denominator for most volatility models. Let's symbolize System Marginal Price, SMP by $P_t$. We take the $\log P_t$ as the only transformation. We treat $P_t$ as a stochastic process exhibiting a degree of correlation from one measurement to the next. We can exploitate this correlation structure found in our data, to predict future values of the process based on the past history of observations. Exploitation of the correlation structure provides us with an opportunity to decomposing $P_t$, into the following components: a **deterministic component** or the **forecast** (a term used from now on), a **random component** (the error or uncertainty, related to the forecast).

So, the general expression of the model for $\log P_t$ is:

$$\log P_t = f(t-1,X) + \varepsilon_t \qquad (4.2)$$

where $f(t-1,X)$ is a nonlinear function representing the **forecast** or deterministic component, of the **current price** as a function of the **information set** available at time $t-1$, $\Psi_{t-1}$. The **forecast** may consists of past disturbances { $\varepsilon_{t-1}$, $\varepsilon_{t-2}$, ....}, past observations { $P_{t-1}$, $P_{t-2}$, ....}, and any other relevant explanatory time series data X (e.g. Load, Temperature, fuel prices etc.). { $\varepsilon_t$} is a random **innovation process**, representing disturbances in the mean of { $P_t$ }, but $\varepsilon_t$ can also be interpreted as the one step ahead or the single-period-ahead forecast error, since from (4.2) we have

$$\varepsilon_t = P_t - f(t-1,X)$$

It is also assumed in this study that the innovations $\varepsilon_t$ are random disturbances with zero mean and uncorrelated from one time step to the other, i.e. $E\{\varepsilon_t\}=0$, $E\{\varepsilon_{t1} \varepsilon_{t2}\}=0$, $t_1 \neq t_2$.

Although $\varepsilon_t$ are uncorrelated they are not independent but successive values depend on each other as in case of a generating rule $\varepsilon_t$

$$\varepsilon_t = \sigma_t z_t \tag{4.3}$$

where $\sigma_t$ is the **conditional standard deviation** and $z_t$ is a standardized, independent, identically-distributed (iid) random sample generated by a Normal or Student's t probability distribution.

Equation (4.3) says that $\{\varepsilon_t\}$ **rescales** an iid process $\{z_t\}$ with a **conditional standard deviation, $\sigma_t$,** that includes the serial dependence of the innovations. This means that the quantity, $\varepsilon_t/\sigma_t$, the **standardized disturbance,** is also iid.

It is well known from finance theory that GARCH models are consistent with the concept of efficient market hypothesis (EMH), according to which **observed** past returns cannot improve the forecasts of future returns (Campbell Y.J., et.al, 1997). Therefore, GARCH innovations { **$\varepsilon_t$** } are serially uncorrelated.

### 4.4.1 Conditional Mean Models

Since we are interested in modeling the impacts of a variety of explanatory variables on SMP, both of a "fundamental" type of relation to SMP (e.g. load, fuel prices, type of generation etc.) and of "external" type (the regulator's changes or policies), we adopt an ARMAX model for the conditional mean, as given in (4.4)

The expression for the nonlinear function f(t-1,X) can take the general form of an ARMAX(R, M, $N_x$) model for the conditional mean:

$$logP_t = c + \sum_{i=t}^{R} \varphi_i logP_t + \varepsilon_t + \sum_{j=1}^{M} \theta_j \varepsilon_{t-j} + \sum_{K=1}^{N_x} \beta_k X(t,k) \tag{4.4}$$

where $\varphi_i$ the autoregressive coefficients, $\theta_j$ the moving average coefficients, $\varepsilon_t$ innovations (or residuals) a zero-mean white noise process with variance $\sigma^2$, $\varepsilon_t \sim WN(0, \sigma^2)$. X(t,k) is an explanatory regression matrix of k variables. R, M are the orders of the Autoregressive and Moving average polynomials and $N_x$ the number of explanatory variables in the matrix X.

### 4.4.2 Conditional Variance Models

The **conditional variance of innovations** is given below:

$$Var_{t-1}(logP_t) = E_{t-1}(\varepsilon_t^2) = \sigma_t^2 \tag{4.5}$$

We must point out here the distinction between the **conditional** and **unconditional** variances of $\{\varepsilon_t\}$. **Conditional** variance implies explicit dependence on a past series of observations. **Unconditional** refers to **long-term** evolution of a time series and assumes no explicit knowledge of past observations.

Following a similar approach with Lo and Wu (Lo K.L. and Y.K. Wu, 2004), let $\varepsilon_t$ a real-valued discrete-time stochastic process, $\psi_{t-1}$ the information set of all information through time t-1, and $\sigma_t^2$ the conditional variance. The general GARCH (P,Q) model for conditional variance of innovation is

$$E_t | \psi_{t-1} \sim N(0, \sigma_t^2) \qquad (4.6a)$$

$$\sigma_t^2 = k + \sum_{i=1}^{P} G_i \sigma_{t-1}^2 + \sum_{j=1}^{Q} A_j \ \varepsilon_{t-j}^2 \qquad (4.6b)$$

where k>0, $G_i \geq 0$ and $A_j \geq 0$

This is a symmetric variance process i.e. it does not take into account the sign of the disturbance.

$G_i$ is a key **persistence parameter** of which high value implies a high impact of past volatility to future volatility, while a low value implies a lower dependence on past volatility. Volatility is said to be persistent if today's return has a large effect on the future or forecasted variance many, say, days ahead. This is equivalent in saying that periods of high and low volatility tend to be grouped or **clustered.** The significance of $G_i$ will be shown in section 10.1.6 below.

When P=Q=1 in (4.6b) we get the GARCH(1,1) model, and for R=M=0 and no explanatory Matrix X, we have a constant mean equation (f(t-1)≡constant) and normally distributed innovations. This simple model has been shown to capture the volatility in most of financial returns (see Lo et. al, 2004 and references in it). We used this model in this work to test if it can do the job in SMP series.

$$\varepsilon_t | \psi_{t-1} \sim N(0, \sigma_t^2) \qquad (4.7a)$$
$$\log P_t = c + \varepsilon_t \qquad (4.7b)$$
$$\sigma_t^2 = k + G_1 \sigma_{t-1}^2 + A_1 \varepsilon_{t-1}^2 \qquad (4.7c)$$

Generally, this model satisfies the stylized facts of persistence (or clustering) and mean reverting. $G_1 + A_1$ is a point estimate **of persistence**, i.e. the time taken for volatility to move halfway back to its unconditional mean value following a given perturbation. If $G_1 + A_1 < 1$ then we have a mean reverting conditional volatility mechanism in which perturbations or shocks are transitory in nature. The closest this parameter is to unity the slower the shocks to SMP volatility die out. The number of days for volatility to return or revert half-way back to its mean is given as $\ln(1/2) / \ln(G_1 + A_1)$. We have calculated this **half-life** parameter as described below. The asymptotic relationship of the unconditional $\sigma^2$ with the conditional one given by (4.7c) is $\sigma^2 = \frac{k}{1 - G_1 - A_1}$.

The $\log P_t$ series, in (4.7), consists of a constant plus an uncorrelated white noise disturbance $\varepsilon_t$. The variance forecast, $\sigma_{t-1}^2$, is the sum of a constant k plus a weighted average of last period's forecast and last period's squared disturbance. As we have seen in section 2, the $\log P_t$, exhibit significant correlation and persistence. Thus, our data for $\log P_t$ is a candidate for GARCH modeling and we have already mentioned that exhibit volatility clustering. We will investigate whether the innovations of the fitted model to the $\log P_t$ have an **asymmetric impact** on the price volatility. It is rational to expect that positive shocks to SMP increase volatility more than negative shocks. This rationality stems from the fact that an increase to SMP (a positive shock) corresponds, naturally, to say, an unexpected increase in demand for electricity. Furthermore, because marginal costs are convex, positive demand perturbations have a larger influence on price changes relative to negative perturbations. The above situation describes the **leverage effect** that we mentioned earlier and is captured in finance by the EGARCH model of Nelson (Nelson D.B., 1991).

One type of GARCH model useful in capturing the **leverage effect** (or asymmetric volatility effect) is the general EGARCH(P,Q) model, for the conditional variance of the innovations.

$$log\sigma_t^2 = k + \sum_{i=1}^{P} G_i \, log\sigma_{t-i}^2 + \sum_{j=1}^{Q} A_j \left[\frac{|\varepsilon_{t-j}|}{\sigma_{t-j}} - E\left\{\frac{|\varepsilon_{t-j}|}{\sigma_{t-j}}\right\}\right] + \sum_{j=1}^{Q} L_j \left[\frac{\varepsilon_{t-j}}{\sigma_{t-j}}\right] \quad (4.8a)$$

where

$E\{|z_{t-j}|\} = E\left[\frac{|\varepsilon_{t-j}|}{\sigma_{t-j}}\right] = \sqrt{\frac{2}{\pi}}$ for normal distribution and $E\{|z_{t-j}|\} = E\left[\frac{|\varepsilon_{t-j}|}{\sigma_{t-j}}\right] = \sqrt{\frac{v-2}{\pi}} \frac{\Gamma\left(\frac{v-1}{2}\right)}{\Gamma\frac{v}{2}}$

for student's t distribution with degrees of freedom v>2. $L_j$ is the leverage term.

For the EGARCH(1,1) the **Conditional Variance** equation becomes

$$log \, \sigma_t^2 = k + G_1 \, log \, \sigma_{t-1}^2 + A_1 \left[\frac{|\varepsilon_{t-1}|}{\sigma_{t-1}} - \sqrt{\frac{2}{\pi}}\right] - L_1 \frac{\varepsilon_{t-1}}{\sigma_{t-1}} \quad (4.8b)$$

where $L_1$ is the **asymmetry parameter.** We note also that the **magnitude effect** is measured by $A_1 \left[\frac{|e_{t-1}|}{\sigma_{t-1}} - \sqrt{\frac{2}{\pi}}\right]$ and the **sign effect** is indicated by $L_1 \frac{\varepsilon_{t-1}}{\sigma_{t-1}}$.

If $L_1 = 0$, then there is no asymmetric effect of past shocks on current variance.
If $-1<L_1<0$ then a positive shock increases variance less than a negative shock.
If $L_1<1$ then positive shocks reduce variance while negative shocks increase variance.

Parameter estimates for (4.8b) were found to be not significant, in our case. For this model (4.8a) the approximate relationship between the unconditional variance σ², of the fitted innovations process, and the $G_1$ parameter is $= \sqrt{\frac{k}{e^{1-G_1}}}$.

Another model that is used for capturing the leverage effect is the general GJR(P,Q) and
For P=Q=1 we have the **GJR(1,1)** with an equation for the **conditional variance** as follows

$$\sigma^2 = k + A_1\varepsilon^2_{t-1} + L_1 S_{t-1}\varepsilon^2_{\tau-1} + G_1\sigma^2_{t-1} \quad (4.8c)$$

where $s_{t-1}$= 1 if $\varepsilon_{t-1}<0$ and 0 if $\varepsilon_{t-1}\geq 0$

The variable $s_{t-1}$ is an indicator variable to account for the effect of positive shocks or good news ($\varepsilon_{t-1}\geq 0$) and the negative shocks or bad news ($\varepsilon_{t-1}<0$) in the market. Therefore, the effect on volatility of good news is $A_1$ while in the case of bad news the effect is $A_1+L_1$. In this model also, K is positive and $A_1$ nonnegative (to ensure a positive variance. The EGARCH and GJR models are asymmetric ones that capture the leverage effect, or negative correlation, between $r_t$ and $\sigma^2_t$. This means that these models explicitly take into account the **sign** and **magnitude** of the $\varepsilon_t$, the innovation noise term. Therefore if the leverage effect does indeed hold, the leverage coefficients $L_j$ should be negative for EGARCH models and positive for GJR. The terms $G_i$ and $A_i$ capture **volatility clustering** in GARCH and GJR models while in EGARCH model volatility is captured only via $G_i$ terms.

## 4.5 Model specifications, outputs and diagnostic tests

We have tested a number of promising (composite) symmetric and asymmetric models in searching for the best one to capture successfully the stylized facts in model's innovations. Table 4.2a show the results, the Akaike and Bayesian Information Criteria (AIC, BIC) and the Durbin-Watson statistic. The model in (4.7) was found to be inappropriate (in replicating the dynamics of SMP) by using a combination of criteria for assessing model's specification. Yet, this model can't be used to detect the effects of fundamental and regulatory factors on SMP.

Model 10 seems to be is the best of all in the set of symmetric and asymmetric models since a combination of AIC and BIC criteria , DW and the ARCH test for detecting any remaining serial correlations in the residuals of the variance equation, are in favor of this model. Durbin – Watson (DW) statistics is used to analyze whether the residuals are clean, i.e. tests the hypothesis that there is no first-order autocorrelation or serial correlation apparent in the residuals. The residuals must be clean and have no structure. If 2<DW<4 then there is negative correlation in the residuals while for DW<2 positive correlation. As a rule of thumb, a DW of exactly 2 indicates clean residuals and any substantial deviation from this value indicates 1st order serial correlation in the residuals. Model 10 in table 4.2a, which is our best found model has DW=2.036 close to the perfect value so we can trust our model of not having first order serial correlations in the residuals. The existence of higher order serial correlation is detecting through ARCH test, and the correlograms of standardized residuals square (Ljung-Box Q-Statistics), as described below.

The specification of our best model 10 is therefore ARMAX(7,1,25)/GRCH(1,1). The complete form of the system of equation of the model is:

**conditional mean** (equation (4.4))

$$\log P_t = c + \varphi_1 \log P_{t-1} + \varphi_2 \log P_{t-2} + ....\varphi_7 \log P_{t-7} + \theta_1 \varepsilon_{t-1} + \beta_1 X(t,1) + \beta_2 X(t,2) + ......+ \beta_{25} X(t,25)$$

(4.9a)

**conditional variance** $\sigma_t^2 = k + A_1 \varepsilon_{t-1}^2 + G_1 \sigma_{t-1}^2$ (4.9b)

The results of fitting the model to our data are given in table 4.2a, where the values of the coefficients are provided.

**Table 4.2a:** Assessment of model specifications by applying AIC ARCH test

| Model | Conditional Volatility equation | Specification | Number of Exogenous Regressors | | R-squared | Durbin-Watson Stat | AIC | HIC | Hannan-Quinn | ARCH-test (for 7 lags) | | ARCH TEST (serial correlation remains in residuals Yes/No) |
|---|---|---|---|---|---|---|---|---|---|---|---|---|
| | | | mean equation | variance equation | | | | | | Stat. | p-value | |
| 0 | symmetric | ARMAX(0,0,0)/GARCH(1,1) Base model | 0 | 0 | -0.097 | 0.189 | 0.409 | 0.417 | 0.412 | 106.208 | 0.0000 | yes |
| 1 | symmetric | ARMAX(1,1,25)/GARCH(1,1) | 25 | 0 | 0.916 | 1.818 | -1.886 | -1.820 | -1.862 | 12.255 | 0.0925 | no |
| 2 | | ARMAX(1,1,25)/GARCH(1,1) | 25 | 8 | 0.852 | 0.780 | -1.415 | -1.338 | -1.387 | 41.441 | 0.0000 | yes |
| 3 | | ARMAX(1,1,25)/GARCH(1,1) | 25 | 25 | 0.853 | 0.792 | -1.432 | -1.320 | -1.392 | 48.093 | 0.0000 | yes |
| 4 | | ARMAX(2,1,25)/GARCH(1,1) | 25 | 25 | 0.917 | 2.068 | -1.966 | -1.848 | -1.924 | 31.912 | 0.0000 | yes |
| 5 | | ARMAX(2,1,25)/GARCH(1,1) | 25 | 8 | 0.919 | 2.058 | -1.932 | -1.847 | -1.901 | 26.674 | 0.0004 | yes |
| 6 | asymmetric | ARMAX(1,1,25)/GJR(1,1)* | 25 | 25 | 0.851 | 0.790 | -1.434 | -1.320 | -1.393 | 47.343 | 0.0000 | yes |
| 7 | | ARMAX(2,1,25)/GJR(1,1)* | 25 | 8 | 0.919 | 2.070 | -1.9327 | -1.847 | -1.901 | 26.712 | 0.0004 | yes |
| 8 | | ARMAX(2,1,25)/EGARCH(1,1)* | 25 | 8 | 0.919 | 2.050 | -1.930 | -1.845 | -1.899 | 30.658 | 0.0001 | yes |
| 9 | symmetric | ARMAX(7,1,25)/GARCH(1,1) | 25 | 25 | 0.918 | 2.074 | -1.966 | -1.835 | -1.910 | 19.600 | 0.0065 | No |
| 10 | | ARMAX(7,1,25)/GARCH(1,1) | 25 | 8 | 0.921 | 2.036 | -1.971 | -1.878 | -1.938 | 13.331 | 0.0644 | no |

**Note*:** leverage effect coefficient found to be non-significant, see table 4.2b

**Table 4.2b:** Asymmetric model coefficient of leverage effect

| Model | L₁ | p-value |
|---|---|---|
| 6 | 0.1268 | 0.1385 |
| 7 | 0.0843 | 0.3046 |
| 8 | -0.0048 | 0.8892 |

**Table 4.3 :** ARMAX(7,1,25)/GARCH(1,1) model parameters

## Mean Equation

| Variable | Coefficient | Coeff. value | Std. Error | z-Statistic | Prob. |
|---|---|---|---|---|---|
| C |  | 0.637868 | 0.449498 | 1.419068 | 0.1559 |
| D(AHELENR) | $\beta_1$ | -0.000071 | 0.000136 | -0.529789 | 0.5963 |
| D(ALIAKENR) | $\beta_2$ | -0.001604 | 0.000741 | 2.163641 | 0.0305 |
| D(ARAXTHENR) | $\beta_3$ | -0.000245 | 0.000862 | -0.284382 | 0.7761 |
| LOG(BRENT) | $\beta_4$ | 0.256842 | 0.077274 | 3.323766 | 0.0009 |
| LOG(HFOG) | $\beta_5$ | 0.005906 | 0.002383 | 2.477875 | 0.0132 |
| LOG(HYDGEN) | $\beta_6$ | -0.052671 | 0.006899 | -7.634190 | 0.0000 |
| LOG(HYDMR) | $\beta_7$ | -0.028504 | 0.005643 | -5.050850 | 0.0000 |
| D(LADONENR) | $\beta_8$ | -0.003870 | 0.002852 | -1.357035 | 0.1748 |
| LOG(LIGNG) | $\beta_9$ | -0.336946 | 0.025724 | -13.09859 | 0.0000 |
| LOG(LOAD) | $\beta_{10}$ | 1.558140 | 0.031401 | 49.62039 | 0.0000 |
| LOG(NATGASG) | $\beta_{11}$ | 0.047442 | 0.007023 | 6.755063 | 0.0000 |
| D(NESTOSENR) | $\beta_{12}$ | 0.002768 | 0.001042 | 2.655939 | 0.0079 |
| D(PLASTENR) | $\beta_{13}$ | -0.005011 | 0.001603 | -3.124855 | 0.0018 |
| LOG(RESGEN) | $\beta_{14}$ | -0.033070 | 0.002008 | -16.47227 | 0.0000 |
| LOG(TOTEXPORTS) | $\beta_{15}$ | 0.023108 | 0.002264 | 10.20810 | 0.0000 |
| LOG(TOTIMPORTS) | $\beta_{16}$ | -0.127141 | 0.010479 | -12.13301 | 0.0000 |
| LOG(UAVAI) | $\beta_{17}$ | -0.714757 | 0.032649 | -21.89196 | 0.0000 |
| RMR1 | $\beta_{18}$ | 0.176995 | 0.094799 | 1.867053 | 0.0619 |
| RMR2 | $\beta_{19}$ | 0.161638 | 0.099196 | 1.629483 | 0.1032 |
| RMR3 | $\beta_{20}$ | 0.234799 | 0.106144 | 2.212081 | 0.0270 |
| RMR4 | $\beta_{21}$ | -0.151792 | 0.025032 | -6.063320 | 0.0000 |
| RMR5 | $\beta_{22}$ | 0.262312 | 0.044900 | 5.842120 | 0.0000 |
| RMR6 | $\beta_{23}$ | -0.364146 | 0.041763 | -8.724973 | 0.0000 |
| RMR7 | $\beta_{24}$ | 0.107159 | 0.045132 | 2.374328 | 0.0176 |
| RMR8 | $\beta_{25}$ | 0.130810 | 0.103499 | 1.63875 | 0.2063 |
| AR(1) | $\varphi_1$ | 0.236197 | 0.140190 | 1.684830 | 0.0920 |
| AR(2) | $\varphi_2$ | 0.203822 | 0.070729 | 2.881717 | 0.0040 |
| AR(3) | $\varphi_3$ | 0.088161 | 0.025150 | 3.505458 | 0.0005 |
| AR(4) | $\varphi_4$ | 0.078068 | 0.023089 | 3.381118 | 0.0007 |
| AR(5) | $\varphi_5$ | 0.087547 | 0.021817 | 4.012843 | 0.0001 |
| AR(6) | $\varphi_6$ | 0.069808 | 0.022847 | 3.055447 | 0.0022 |
| AR(7) | $\varphi_7$ | 0.139523 | 0.020427 | 6.830459 | 0.0000 |
| MA(1) | $\theta_1$ | 0.241577 | 0.141563 | 1.706500 | 0.0879 |

**Variance Equation**

|   |   |   |   |   |   |
|---|---|---|---|---|---|
| K |   | 0.001466 | 0.000330 | 4.440105 | 0.0000 |
| RESID(-1)^2 | $A_1$ | 0.134847 | 0.020387 | 6.614217 | 0.0000 |
| GARCH(-1) | $G_1$ | 0.787272 | 0.026990 | 29.16894 | 0.0000 |
| RMR1 |   | 0.000512 | 0.000530 | 0.096472 | 0.9231 |
| RMR2 |   | 0.000942 | 0.003153 | 0.298736 | 0.7651 |
| RMR3 |   | -0.002166 | 0.003153 | -0.690960 | 0.4896 |
| RMR4 |   | 8.27E-06 | 7.21E-05 | 0.114690 | 0.9087 |
| RMR5 |   | 0.000237 | 0.000135 | 1.755964 | 0.0791 |
| RMR6 |   | 0.000411 | 0.000199 | 2.063805 | 0.0390 |
| RMR7 |   | 0.000356 | 0.000279 | 1.275853 | 0.2020 |
| RMR8 |   | 0.000619 | 0.000823 | 0.752420 | 0.4518 |

| | | | |
|---|---|---|---|
| R-squared | 0.921336 | Mean dependent var | 3.977355 |
| Adjusted R-squared | 0.920421 | S.D. dependent var | 0.374881 |
| S.E. of regression | 0.105753 | Akaike info criterion | -1.971799 |
| Sum squared resid | 31.70594 | Schwarz criterion | -1.878290 |
| Log likelihood | 2873.546 | Hannan-Quinn criter. | -1.938088 |
| Durbin-Watson stat | 2.036971 | | |

Therefore, the **conditional variance** equation becomes**:**

$$\sigma^2_t = 0.0011 + 0.134\varepsilon^2_{t-1} + 0.787\sigma^2_{t-1} \qquad (4.10)$$

If we interpret $G_1$ and $A_1$ as weights, $A_1$ indicates that a weight of 13.4% is assigned to the most recent observation. (4.10) clearly shows that volatility is time varying, responding to the most recent data value of SMP. A weight of 79% is attached to the volatility estimate of yesterday.

The **persistence parameter** is given by $P = A_1 + G_1 = 0.922 < 1$ i.e. the model is **covariance stationary. This parameter shows also the existence of a volatility clustering or mean reverting behavior of the SMPep.** The **Half-life parameter** is H=ln(1/2)/ln($G_1$+$A_1$)=9 days, meaning that it takes on average 9 days for half of the volatility shock to fade out.

We have to mention here that a formal **leverage effect** as opposed to **inverse** leverage effect as Knittel and Roberts (2005) and Karakatsani and Bunn (Karakatsani N., et.al., 2004) have reported in their work, where a positive shock will increase the volatility more than a negative one. Our finding is that we do not have any leverage effect since the corresponding coefficients L in the asymmetric model fitting were found to be not significant. The result, however, reported in the work of Gianfreda (2010) for French, German, Italian and Spanish markets, using an EGARCH(1,1) model, and in the work of Montero et. al (2011) has shown the existence of such leverage effects.

The final value of the conditional st.dev or volatility $\sigma_t$ (see fig.4.8), is 0.133. We compare this value to the **unconditional standard deviation** $\sigma$, estimated by the model's parameters as follows:

$$\sigma = \sqrt{\frac{k}{1 - G_1 - A_1}} = \sqrt{\frac{0.0014}{1 - 0.787 - 0.134}} = 0.133$$

Therefore, the **long-term behavior** of $\sigma_t$ forecasts tends to the unconditional volatility of 0.133 given by using the parameters of the estimated model.

Figure 4.6 shows the actual, fitted and residuals (innovations) values of model (4.10) while figure 4.7 depicts the Conditional Variance generated by the same model. On the same figure we provide the dates the regulatory reforms took place, so we can see the spikes or shocks they have induced on the conditional volatility of SMP, as the coefficients in mean and variance equations associated with these dummy variables (reforms) reveal in table 4.3.

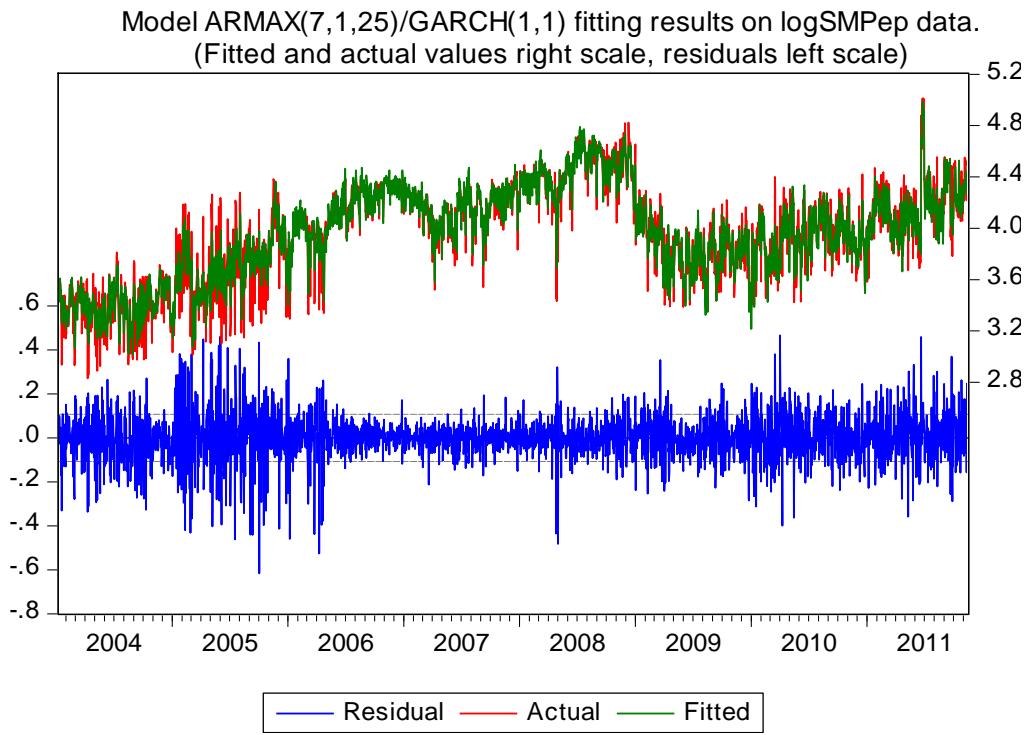

**Figure 4.6 :** ARMAX(7,1,25)/GARCH(1,1) fitting results. Plot of actual, fitted and residuals (innovations).

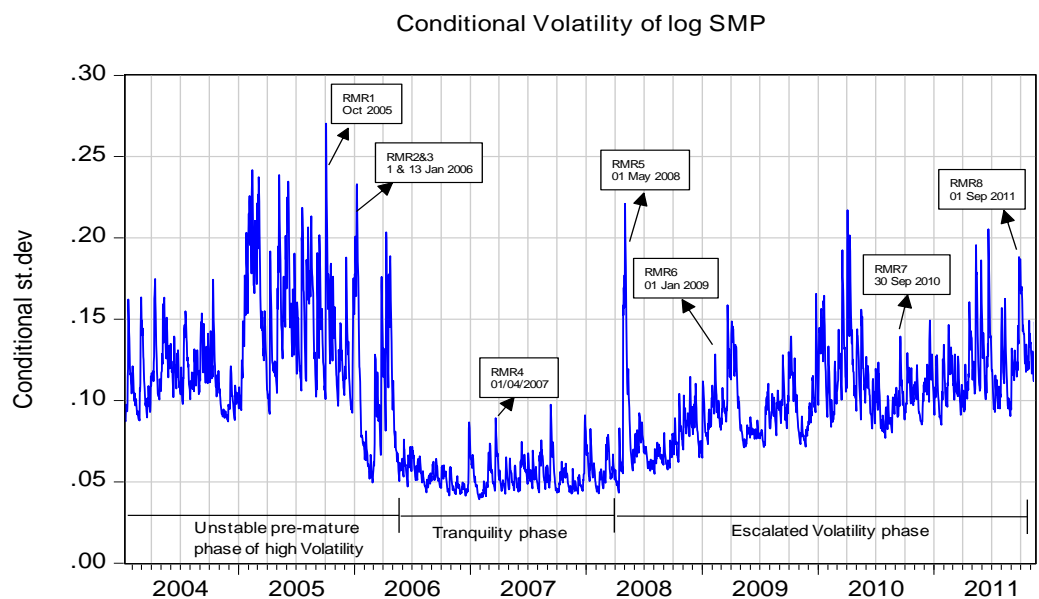

Figure 4.7 Phases of evolution of Conditional volatility of log SMP and dates of RMRs.

The most tranquil phase of the market, in terms of volatility, is from May 2006 to April 2008.

This calmness, however, is broken due to (among other factors) reform 4, generating the second significant spike in this period. The end of this tranquillity comes when reform RMR5 is launched in May 2008, responsible for a high spike taking the volatility from a value of 5% to 22.5% , to drop again to the previous value after a month, passing first through the unconditional value of σ=13% (the long-term mean value), the volatility Shock having lost half of its intensity in almost 10 days (half-life was estimated to be 10 days). After the launching of RMR5, a phase of upward systematic escalation of the conditional volatility begins, with relatively small (compared to the previous spikes) spikes due to reforms 6 and 7. RMR8 gives the last spike shown but this can't be attributed to this reform due to the fact that the coefficient found is insignificant. In this escalation period the conditional volatility is increased from 5% to 13.5% (170%).

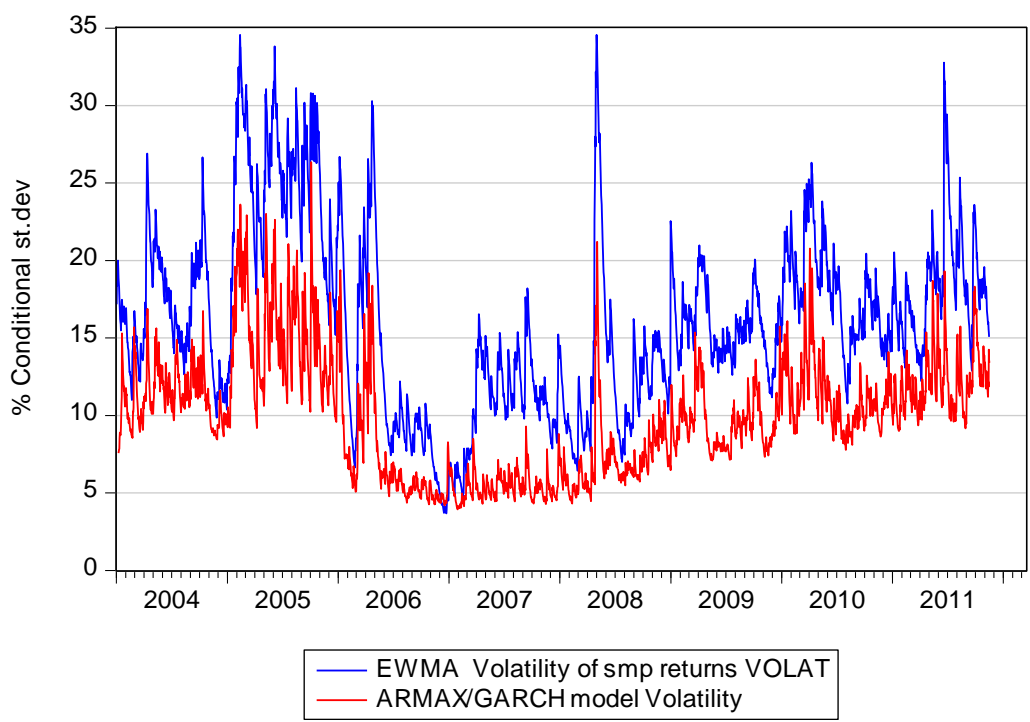

**Figure 4.8:** % Comparing dynamics of Conditional Volatility estimated by ARMAX(7,1,25)/GARCH(1,1) and EWMA models.

Figure 4.8 shows the comparison of the forecasted conditional volatility of the GARCH model, versus the conditional volatility estimated EWMA. This comparison serves as an indication of the model's ability to track the variations in the GEM's volatility. The asymptotic behaviors of the two models forecasts approach a long-term value of 13.4% (the last value red curve) while and 15.3% for the blue curve. Both models exhibit almost the same dynamic behavior , however EWMA model show instantaneous values always larger than GARCH due to differences in $A_1$ (GARCH) and $\lambda-1= A_1$ (for EWMA) parameters.

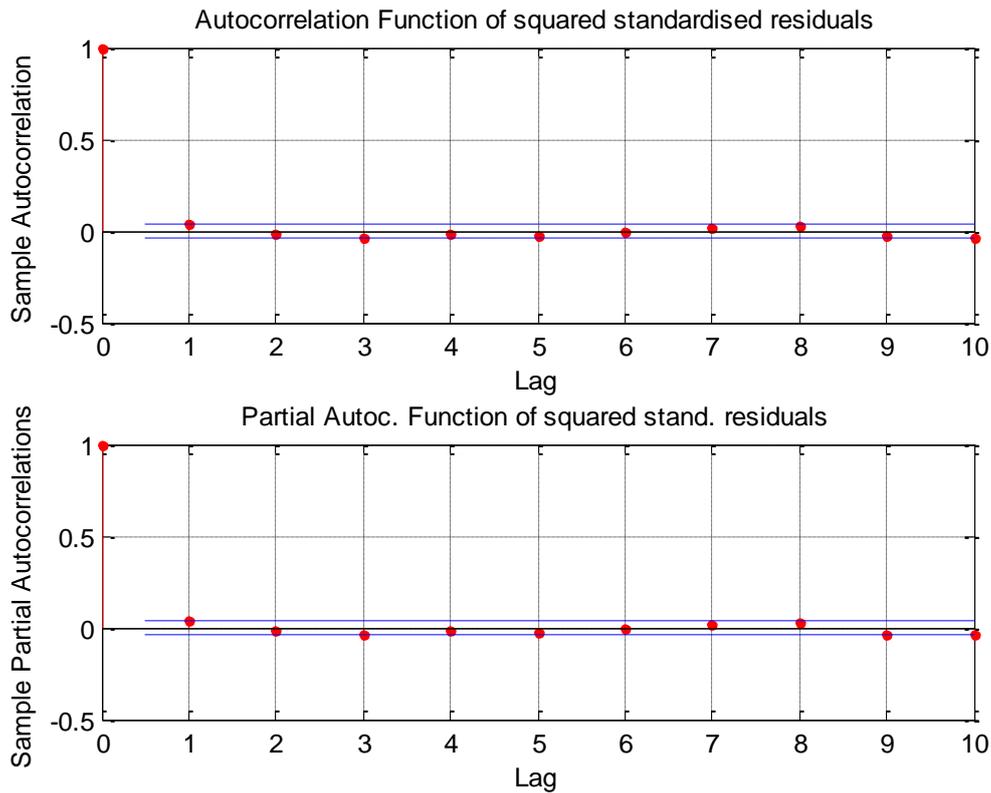

**Figure 4.8** Autocorrelation and Partial autocorr. Functions Of standardized squared Residuals of model (4.9).

Any successful GARCH model is that one in which there is no autocorrelation in the squared standardized returns $\varepsilon_t^2/\sigma_t^2$, meaning that the model is capable of capturing **volatility clustering** in the original returns. Volatility clustering implies a strong autocorrelation in squared returns.

From fig. 4.8 we see that the standardized squared residuals show no correlations (compare with ACF and PACF of SMP in fig. 3.6 before fitting the model). Therefore, our fitted model sufficiently explains heteroskedasticity behavior in our data.

Furthermore, we repeated ARCH-test on these innovations (Table 4.4) and compared the results with those in the pre-estimation analysis. In the pre-estimation analysis the test indicated the existence of both correlations and heteroskedasticity. In the post-estimation analysis, ARCH-test now indicates acceptance of the respective null hypothesis that no ARCH effect exists, up to lags seven. This result further enhances the explanatory power of the fitted model. The ARCH test statistics was found to be 13.33 and insignificant (p=0.0643) for seven lags (all coefficients at lags 1 to 7 are insignificant. So no ARCH effects exist in the innovations of the fitted model.

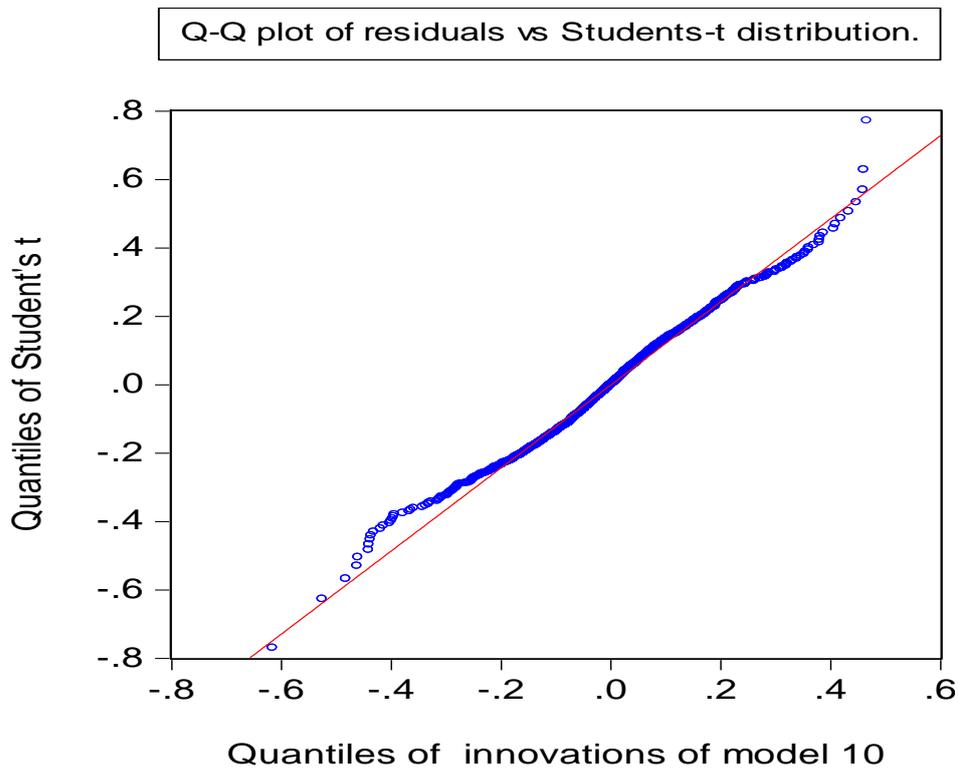

**Figure 4.9** Q-Q plot of model's residuals vs the Student's-t distribution.

Figure 4.9 shows the result of fitting the residuals generated by the model on the theoretical Student's-t distribution that we have chosen to be the distribution of errors in specifying the GARCH(1,1) model. We see a very good fit of the residuals since the largest amount of data are on the theoretical line, except the extreme right, positive values that deviate from the line, forming a **fat tail** of the distribution of the residuals. This fitting of good quality is also an extra evidence that our model has been correctly specified. Using now our best model (4.9,4.10) we perform in-sample or static SMP forecasts as shown in figure 4.10, the dynamics of which mimics very well the dynamics of the actual series in fig. 3.3. The quality of the fit is also very good: Theil's coefficient 0.045, Bias (mean) and variance proportions small (0.003 and 0.022 respectively). The descriptive statistics of the forecasted series are very close to the ones of the actual series (mean 57, max 147, min 20.3, st.dev 19.51, skewness 0.52,kurtosis 3.12), by comparing with the values in table3.3.

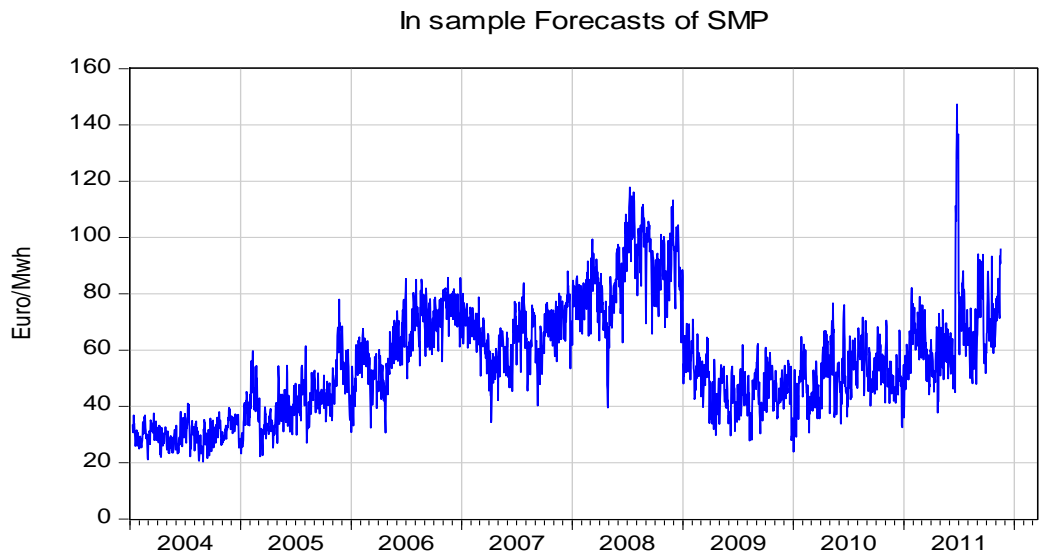

Figure 4.10 In-sample forecasts of SMP ex-post from best model

In figures 7.2 and 7.3 we show the forecasting results of the model corresponding to two different scenarios for the evolution of the conditional variance with different starting times : the first in the tranquil, of low volatility phase, just before the application of RMR5 (end of April 2008) , while the second in a high volatility regime, namely on 4th May 2008 three days after the enactment of RMR5. In the first case the long-term stable value of unconditional variance is approached from below while in the second case from above. These two scenarios confirm the quality of the model in generating rational and consistent dynamic behavior of the conditional volatility.

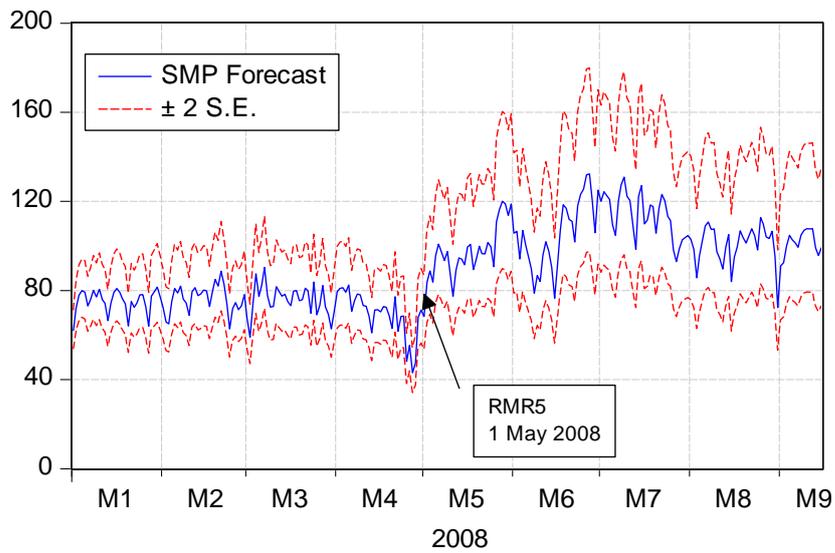

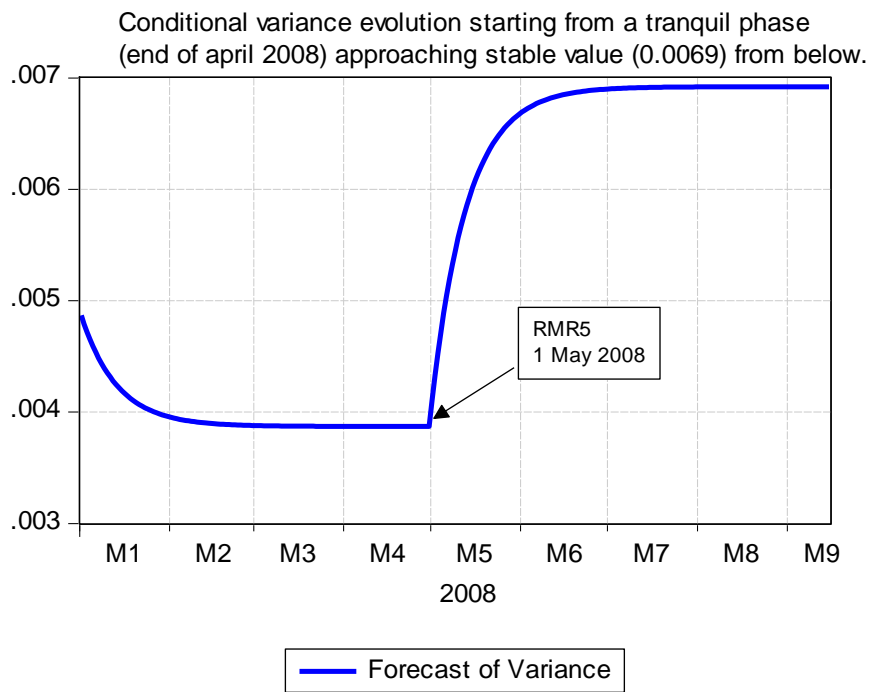

Figure 4.11 Conditional Variance evolution of SMP from end of April 2008 to end of September 2008.

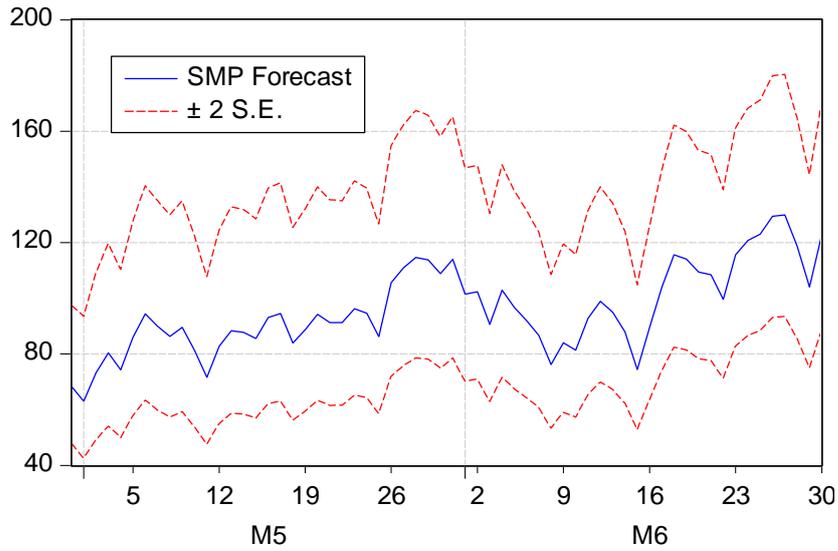

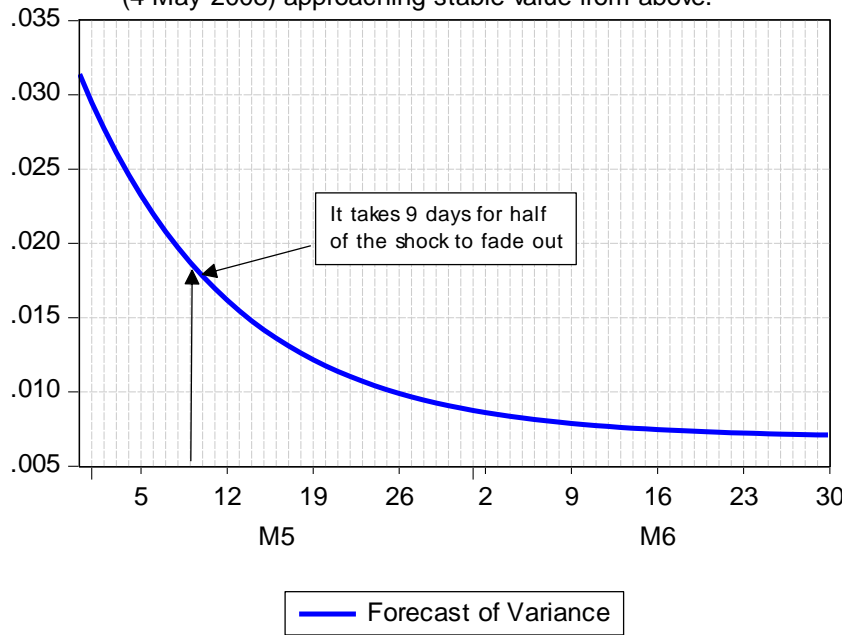

Figure 4.12 Conditional variance evolution of SMP from 4th May 2008 to 30 June 2008.

## 5. Interpreting the results

### The mean equation

Referring to table 4.2b with parameter estimates of the Best Specified model 10, suggests that our time series model confirms the sign and intensity of the effects of fundamental variables on SMP as it is shown in a typical fundamental analysis.

All the quantitative fundamental variables, the autoregressive and moving average structures are robust to controlling for regulatory changes in the GEM's structure. Specifically, both the signs and the values of the coefficients of the fundamental factors remain unchanged before and after the inclusion of RMRs in the model specification.

we observe that hydro generation, both for must-run and for "normal" functioning, have a negative and statistically significant impact on the evolution of SMP. The effect however of the **Hydgen** factor is stronger than that of the **hydmr** (hydro must-run), as the values of the estimated coefficient indicate. This is a logical result as the amount of hydro must-run generated energy is a small part and is included in the series of the total hydro generation (hydgen), used more intensively during peak-hour according to DAS (day-ahead schedule).

The scarcity of hydro generation in 2011 is, among other fundamental factors, responsible for the SMP increase in this year, reflecting the strong and negative correlation of **hydrog** or **hydromr** with SMP. Years 2009 and 2010 instead, were intensely wet, causing a downward force on SMP. However, the driest years were 2007 and 2008 exerting a significant negative 'pressure' on SMP.

The significantly reduced hydro generation in 2011 exerted an upward trend on SMP, mainly due to the need for substitution by more expensive energy (see the negative correlation coefficients of **hydgen** with the thermal generation factors, **brent, hfog, ligng,** and **natgas**).

How much significant is the factor **Hydrogen** in the formation of SMP (through the level and allocation profile of hydro generation), is also shown from its contribution to the formation of SMP's "by-product", the ***retail margin***. A conservative, for example, water management i.e. over-restrictive in particular time periods, can produce a reduction in retail margin, favoring in this way a market supplier that happened to have a reduced retail volume. Therefore a supplier in the retail market with large share and an ability to control hydro generation, as PPC in Greece, can have a crucial impact not only on SMP formation but also, in consequence, on the retail price.

The control over hydro generation factors can be done by designing a new approach of managing hydro stations by considering aspects such as the opportunity cost of water, the cost of fuel mix substituted by water and by a close linking of the above factors to the level of reservoirs.

A negative effect on SMP have also the production of RES (**resgen**) and Lignite (**lignitegen**) stations with different relative strengths as the values of their corresponding coefficients reveal. A careful examination of the evolution of the share of each type of generation in the total annual installed capacity as well as in the duel mix generation, as it is shown in Tables 2.1 and 2.2, reveals that RES share generation is always less than that of hydro as well as Lignite stations over the period 2004-2011, for each year.

The higher the production by lignite stations the lower the SMP, reflecting the impact of a low price input factor to the generating portfolio of the system. Negative, also, effect on SMP have the total emported energy (**totimports**) and the generating unit availability (**Uavail**). The factor has the next higher coefficient but now with a negative impact on SMP is Unit availability for (**uavail**) since an 1% increase in this variable, ceteris paribus, reduces 0.71% the value of SMP. Planned and unplanned outages for maintenance reduce availability. Unplanned outages are stochastic events distributed uniformly over the year. Efforts are made so planned outages are scheduled to happen during periods of low load and low SMP.  The more available are the generating units the lower the SMP, since enough of the required capacity is readily available and according to the DA and dispatching schedules (the units have hardly competing each other to be included on the list for next day's generation). More specifically Unit availability (**uavail**) depends on the level of outages and maintenance and is negatively related to SMP. For example, in the end of 2011 two severe outages in IPP's plants and one in a Unit of PPC (Lavrio5) induced these **Units unavailable** for long periods of time.  The induced, due to this fact cuts in demand, exerted an upward pressure on SMP. During June of the same year a labor Union strike at PPC caused a major capacity withholding i.e. forced power cuts, reducing availability in 20 generating Units which in turn had an extremely strong upward escalation of SMP (the price escalated from 42 €/MW in 19.06.2011 to the price cap (an administratively set price)  of 150 €/MW in 24.06.2011 and finally to 148€/MW in 29.06.2011, being on average at 139.7 €/MWh for these ten days).

Positive effect on SMP has the 3-month moving average of Brent Oil price (Brent) and the generation of Stations that use Heavy fuel Oil (**Hfog**), an expected result. The higher the values of these two variables, the higher the SMP. The energy production of Natural Gas (**natgasg**) Stations also have a positive influence on the behavior of SMP, as this type of fuel is an expensive production input.  The extent of natural  gas's  impact  on SMP is reasonable , taking into consideration the share of this fuel in the generation mix (see tables 2.1 and 2.2) . Yet, the positive coefficient reflects the strong correlation between SMP and Natural Gas prices and more precisely corresponds to the SMP-GAS prices **spark spread**[8], defined here as the difference between SMP and daily Balancing price of gas (see figure 5.1).  The escalation of spark spread volatility to increasing values is evident after the second quarter of 2010, exerting a positive effect on SMP.  The reforms in the market may have broken this positive co-variation, a fact difficult to be captured by the model by only considering changes in the levels of these two quantities.  Figure 5.2 shows the percent change (annual rate) of monthly SMP versus Natural Gas prices. The spikes in spark spread, SMP ex-post and SMP Das observed in the quarters shown in the graph reflect the strong positive co-movement of these series.

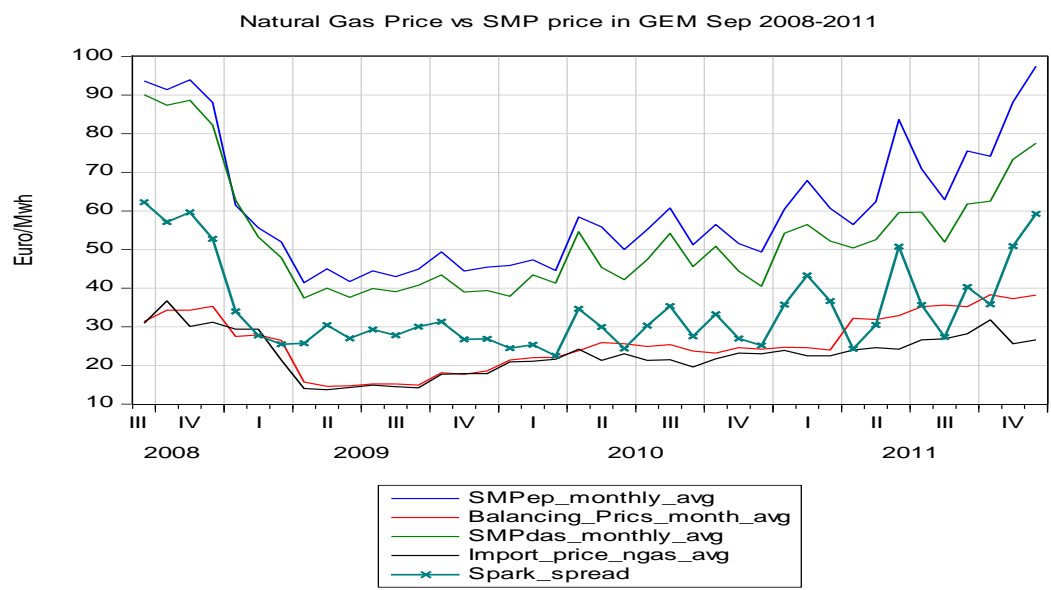

---

[8] We have used monthly weighted-average import price (WAIP) and the daily balancing gas (HTAE) for the same months, given in DESFA's internet site from Jan 2009 to Dec 2011. The data are also available in RAE's website. As spark spread here we mean the difference of SMP and Balancing price of Nat gas.

Figure 5.1 Monthly average Nat Gas price versus SMP in GEM (Sep. 2008-11).

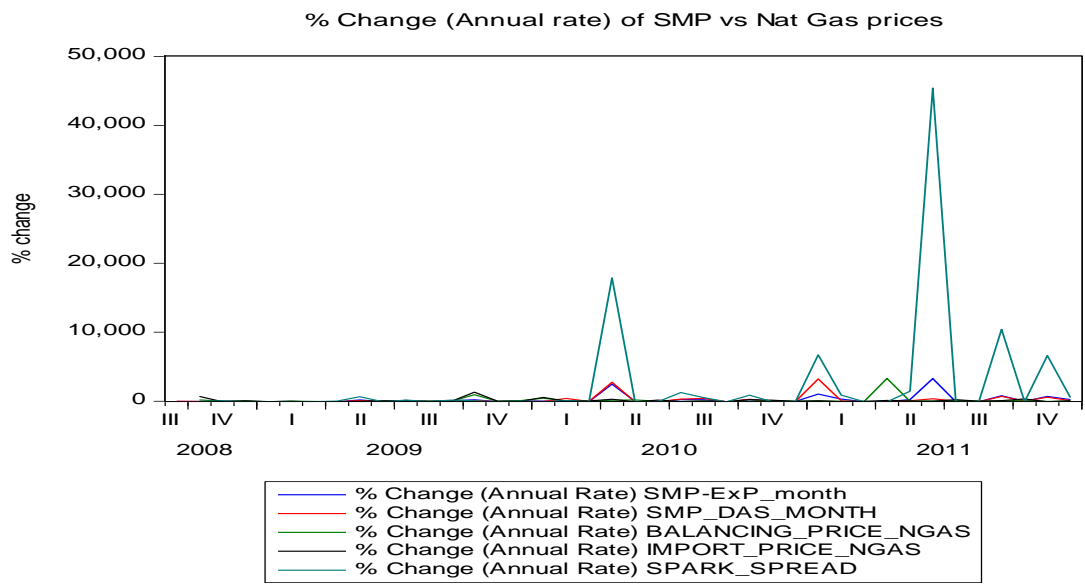

Figure 5.2 % change (annual rate) of SMP and Nat Gas prices in GEM (Sep. 2008-11)

The total export of energy (**totexports**) increase SMP as expected, a rational outcome as energy extracted from the system reduces the supply for domestic usage. The heaviest positive impact on SMP comes from **Load** i.e. the total demand of all customers of the system. The larger the demand (i.e. the further shift to the right of the demand curve) the much higher the new equilibrium point i.e. the point of intersection of Supply and Demand Curves. The coefficient corresponding to load has the largest value indicating that SMP is mostly driven by this variable, as it is well known also through models based on fundamental analysis. So, an increase of 1% in the daily demand, over the period of our analysis, leads to a multiple increase, ceteris paribus, in the level of SMP of 1.56%. An example enhances the view of this strong positive dependence. In November 2011, **load** (demand) increased of about 8% compare to the previous year same month, due to an increased usage of air-conditioning for heating (oil was considered a more expensive alternative to electricity for households), causing an upward pressure to SMP ( i.e an 8%x1.56%=12.5% increase in SMP).

An increase of 1% in the **Brent Oil price** will result, on the average, in an increase of 0.25% while a similar increase in the level of production by a station using **heavy fuel oil** is accompanied by a positive and negligible increase in SMP (0.006%). This result is in accordance with the relatively very small share (on the order of 5% to 6% each year) of this type of generation, as shown on Table 2.1.

The need for linking hydro generation patterns to reservoir levels is further enforced by considering their strong correlation with the corresponding energy reserves that are strongly and negatively correlated with SMP (see correlations table in the Appendix 4).

A combination of factors occurring concurrently can have a significant impact on SMP, even reducing it to almost zero level. For example, a significant drop in **load (demand)**, as in Easter holiday, combined with compulsory generation quantities due to renewables, Plant's technical minima and **imports**, can create an imbalance between production and consumption (an excess generated quantity in our case). Due to this excess, imports can be offered at a zero value and it is necessary to impose a minimum value (curtailed value) in this extreme situation. In 2009 we have witnessed such an extreme case.

## 6. The Impacts of RMRs on SMP

The main message coming out the table 4.3 is that the regulatory interventions in the market architecture have significantly affected the conditional mean of the SMP price and partially its conditional volatility (only RMR6 at 5% level is significant). The results suggest that the average SMP was affected by the regulatory reforms, *ceteris paribus*, with upward, downward trends and spikes attributed to the signs and intensities of the found coefficients.

In the mean equation, the dummy variables representing the Regulatory Market Reforms, RMR, were found to have a significant (most at a 1% level) effect on SMPep, except RM2 and RMR8 that found to have no impact. Positive impact have RMR1, RMR3, RMR5 and RMR7 and negative impact RMR4 and RMR6. The highest positive impact on SMP is due to market reformation RMR5 (value of coefficient 0.26) regarding the **Cost Recovery mechanism, CRM** (see page 10 and the reference mentioned there). The next largest positive impact (0.23) comes from **RMR3** reform, regarding the first change of methodology of calculating system marginal price (SMP).

It is pointed out here that this reform was not expected by the Regulator to have any significant effect on the evolution of SMP (RAE 2010, and Kalantzis F., et.al, 2012). A very crucial point here is that, in general, there are motives of Generators for a systematic abuse of CRM (in case this mechanism is in place) that allows them to bidding below their marginal cost. This strategy of Generators became possible after the launching of RMR6 which in combination with CRM gave the Generators the opportunity to make extra profits, when SMP exceeded their cost (this market rule allows a generator to offer 30% of its capacity at a price less than its minimum marginal cost, while at the same time they can reap a retained compensation at this cost plus 10%). Therefore, the overall, combined, effects of RMR5 and RMR6 have a rather distorting effect on the market. In the time of writing this paper, the authors were aware of RAE´s intentions to propose measures to strongly reduce this distortion by changing or even removing CRM. It turned out that RMR6 is a critical exogenous (relative to fundamental factors) qualitative factor in shaping the dynamics of SMP. Even though this reform allows the dispatch of various Units for providing reserves to the system (a crucial factor in securing Unit´s viability in a period of excessive capacity), it heavily suppresses SMP to values not reflecting generation cost. The negative effects of RMR5 and RMR6 on SMP have also crated two side effects. First, due to suppressed SMP, and given the feed-in-tariffs mechanism on which RES producers are compensated on a special levy level, the need for this purpose cash outflows increased dramatically and secondly the creation of SMP values not corresponding to the ones expected by the fundamental underlying operating variables of the system that governs SMP.

As far as the impact of RMR8 on SMP is concerned, coefficient $\beta_{25}$ in table 4.3 was found to be no significant. The Regulator however had expected an increase to SMP due to this reform. This is due to the fact that the length of the data set after the date of enactment of this reform, Sept. 2011, is very small, therefore the effect on SMP of this dummy variable, on average, is insignificant. However, having a look at the results of year 2011, we see that this tax, which induced asymmetric effects on gas-fired electricity generation versus lignite and imports (actually the variable cost of the gas plants increased 5.4 €/MWh), overall has induced a rise in SMP. During the same year, 2011, the combination of this reform with a decreased Hydro generation (represented by variable **Hydrogen** in the model) had, as a result, a sustainable SMP rise, from the date of application of RMR8 onwards. More specifically, the daily average ex-post SMP increased from 63.71 €/MWh on 31st August 2011 to 90.75 €/MWh on 1st September (+ 42% ). In general, the average SMP during 2011 before RM8 was 65.00 €/MWh while after the reform changed to 83.86 €/MWh. The average price in 2011 was 71.70 €/MWh compared to 52.25 €/MWh in 2010 (see table 3.3).

Comparable impact to the one by RMR3 has the reform RMR1 (0.177) which corresponds to adoption of the mandatory pool (ex-post settlement) as the preferred electricity market model in Greece. This reform that changed the methodology of estimating SMP, in combination with RMR3, seems to have generated not an "additive" but a "multiplicative" effect. One natural explanation is that before the enactment of those two reforms, SMP was determined mainly by Lignite generating Units having an average marginal cost of about 27€/MWh, which offer the implementation of RMR3 the same units determine SMP only during the period off-peak hours, while at the same time Natural Gas and Oil Units (with a marginal cost of around 60€/MWh) started to set SMP for largest time share during peak hours (almost 60% of total hours).

If we multiply the difference in marginal cost (60€/MWh - 27€/MWh = 33€/MWh) with the time share 60% mentioned above we have a net effect of 33€/MWhX0.60≅20€/MWh.

**Table 6.1:** Impacts of Regulatory Reforms on SMPep

| Market Reform with significant impact | Coefficient | Value | % impact on SMPep* |
|---|---|---|---|
| RMR1 | $\beta_{18}$ | 0.1769 | 15.35 |
| RMR3 | $\beta_{20}$ | 0.2347 | 25.86 |
| RMR4 | $\beta_{21}$ | -0.1517 | -14.00 |
| RMR5 | $\beta_{22}$ | 0.2623 | 29.60 |
| RMR6 | $\beta_{23}$ | -0.3641 | -30.50 |
| RMR7 | $\beta_{24}$ | 0.1071 | 11.30 |

**\*Note:** the % impact of a $RMR_v$ dummy variable (changing for 0 to 1), ceteris paribus, having a coefficient $\beta_v$ is $100(e^{\beta_v} - 1)$.

Table 6.1 gives the impact of RMRs on SMP. For example, the SMP (ex post) value on the day of application of RMR3 (January 13, 2006) was 41.04€/MWh. On next day, the price went to 49.36€/MWh i.e. an increase of 8.32€/MWh (20.3%), on January 15 went to 48.65€/MWh i.e. an 18.54% increase and three days later, on January 16, the price increased to 52.05€/MWh, i.e. 26.82% close to 25.86%, as given in the table above. The market realized the impact (due to its inertia) after 3 days.

Based on the above remarks and the entries in table 2.5, we observe that our model revealed different effects of RMR6, RMR7 on SMP, than the Regulator's expected ones. Should this model had been available to the regulator before the launching of the reforms, it would be very useful in providing a better picture on the expected effects of those reforms as well as on RMR4 and RMR5.

<p align="center"><strong>The conditional variance equation</strong></p>

From table 4.3 we see that the quantified effects of the reforms on the volatility of SMP are negligible, except reform RMR6 which is significant at 5% level, however with negligible magnitude (0.0004%). However, we must point out here that since the reforms affect significant the average SMP returns and that the two equations are coupled (see system of equations 6.10), these reforms affect volatility indirectly.

## 7. Comparing conditional volatility of SMP in GEM with SMP in other Markets

**Table 7.1** shows the parameter estimates of the AR(1)/GARCH(1,1) model on SMP, applied in various electricity markets taken from the work of Escribano et al. (2002), amended by the results found in this work for the Greek electricity market and the work of Petrella (2012) for the Italian market. So, we can compare the rate or speed of mean-reversion of the Greek market with those in the paper. The factor $A_1+G_1$ expresses the **persistence** parameter of the **conditional volatilities** to its mean value. In Financial markets it is common that G is larger than 0.7 but A tend to be less than 0.25. If *A+G is close to one* then the persistence is *high* and a shock in the series will decay slowly. A **low** persistence value i.e **A+G<<1** leads to the *fast decay* of the shock to its long-run volatility. If A+G=1 we have a non-stationary or non-stationary or Integrated GARCH model(I-GARCH) for which term structure forecasts do not converge (there is **no mean reversion**)

**Table 7.1:** Variance equation parameters of GARCH(1,1) model applied in various markets
(source : Alexander C., 1998, Escribano et al., 2002, Petrella et al., 2012)

| **Financial Markets** | $A_1$ | $G_1$ | $A_1+G_1$ | Half –life (days)[9] |
|---|---|---|---|---|
| Equities | | | | |
| UK | 0.105 | 0.810 | 0.915 | 11 |
| GE | 0.188 | 0.712 | 0.900 | 7 |
| US | 0.271 | 0.641 | 0.912 | 8 |
| JP | 0.049 | 0.935 | 0.984 | 43 |
| NL | 0.146 | 0.829 | 0.975 | 28 |
| USD rates | | | | |
| DEM | 0.113 | 0.747 | 0.860 | 5 |
| JPY | 0.102 | 0.763 | 0.865 | 5 |
| GBP | 0.028 | 0.935 | 0.963 | 19 |
| NLG | 0.125 | 0.735 | 0.860 | 5 |
| ESP | 0.160 | 0.597 | 0.757 | 3 |
| AUD | 0.241 | 0.674 | 0.915 | 8 |
| Index | | | | |
| ASE (Greece) | 0.100 | 0.900 | 1.00 | Undefined |
| **Electricity Markets** | | | | |
| Greece | 0.13 | 0.79 | 0.92 | 9 |
| Italy | 0.47 | 0.576 | 1.04 | Undefined |
| Spain | 0.18 | 0.78 | 0.96 | 17 |
| Nord Pool | 0.41 | 0.59 | 1.00 | Undefined |
| Argentina | 0.85 | 0.37 | 1.22 | Undefined |
| Australia (Victoria) | 1.07 | 0.49 | 1.56 | Undefined |
| New Zeeland | 0.40 | 0.60 | 1.00 | Undefined |
| PJM | 1.11 | 0.49 | 1.61 | Undefined |

In the markets of **table 7.2**, Nord Pool, Argentina, Australia, New Zeeland, Italy and PJM have persistence parameter equal to one so there is no mean reversion. A low degree of mean-reversion can be explained, according to Escribano, (2002), by the fact that in some of the markets the proportion of electricity generated by hydro resources is extremely large, as in Nord Pool (in the Greek market hydro share accounts for 24% of net installed capacity in 2008-2009, table 2.1). The expected inter-temporal substitution between inputs is more intense in the markets with large amount of Hydro-generation than in the markets with low one, indicating that hydro stations (reservoirs) play the role of the indirect storage of electricity. In the other markets of the table we observe smaller persistence values indicating a higher degree of mean-reversion since generators cannot use inventories to smooth out shocks. It takes nine days in Greece and seventeen in Spain for the conditional volatility to return to their long-run mean value. GEM in this respect behaves like the first three equities market while Spanish market like the USD/GBP rates. We also point out here that mean-reversion in load (demand) or in temperature are the main factors of causing the mean-reversion in electricity prices.

---

[9] Half-life is defined as $\log(0.5)/\log(A_1+G_1)$

Wolak (1997) and Lucia and Schwartz (2002) have also shown that SMP are more "turbulent" or volatile in markets dominated by hydroelectric power. Large dry periods result in significant lower hydro generation, causing in turn an increase, on the average, in SMP. In general, instability of weather conditions have a significant effect on the mean-reversion process (actually its dynamics resembles a unit root process with autoregressive conditional heteroskedastic errors).

The **oligopoly microstructure** of Greek electricity market (at least at its early stage) and the particular rules in place for compensating the Generators, during the transitional period from less oligopolistic to more competitive, are the main factors in the evolution of this different market, compared to "mature" markets e.g. NordPool, Australia etc. (Escribano et al., 2002).

The way stranded costs are treated in a market creates conflictive incentives on the few participating players (Federico G. and A. Whitmore, 1999). The interaction of incentives has direct effects on energy price risk. There is a negative dependence between **stranded costs** to be paid to generators and the price established in the pool. Actually, generators earn higher profits from their participation in the pool, in the case that the prices are higher than a known and pre-established level, but they will get a lower amount of standard costs. These conflicting interests among generators (which depend on the market's microstructure i.e. expectations on the likelihood of receiving stranded costs, generator's market share etc.) have a predictable impact on the equilibrium price i.e. SMP.

**Table 7.2 % Conditional Volatilities of log SMP and ASE Index**

|        | SMP log returns       |             | ASE log returns       |             |
|--------|-----------------------|-------------|-----------------------|-------------|
|        | EWMA (λ=0.94)         | ARMAX/GARCH | EWMA (λ=0.94)         | ARMAX/GARCH |
| max    | 34.60                 | 26.30       | 6.46                  | 5.75        |
| min    | 3.66                  | 3.92        | 0.31                  | 0.62        |
| mean   | 16.12                 | 9.31        | 1.60                  | 1.70        |
| St.dev | 5.95                  | 4.15        | 0.91                  | 0.86        |

## 8. Conclusions

In the beginning of this work we stated its targets, briefly the detection of possible effects of the regulatory changes that have been made in the structure of GEM, on the dynamic evolution of SMP as it is reflected by its statistical properties or stylized effects. We have found that GEM has the majority of stylized characteristics that other electricity markets exhibit and how these compare with those in financial markets. We have given an in depth explanation of the sign and the intensity of the impacts of both fundamental to the GEM factors as well as of the qualitative or dummy variables constructed to capture the regulatory changes. We strongly believe that our findings will contribute to efforts of the authorities entitled to make GEM more efficient and competitive. Our results will be useful in better understanding of the factors of the underline dynamics of SMP, an information needed both in restructuring GEM as well as in better developing the forward market a requirement for compliance of GEM with the Target model.

We have not detected any ***leverage effects*** in GEM as other workers have found in other markets (Karakatsani N., and Bunn,D.W., 2010), Bowden and Payne ,2008 , Knittel and Roberts, 2005 , Hadsell et al.,2004 ). Is this result explained by a lower degree of convexity of the supply curve in GEM compared to higher degree of convexity in other markets, as Kanamura (Kanamura , 2009) suggests?

Our findings enhances the results of Petrella et al. (2012) for the Italian market. We agree with their remark that if an adequate number of similar works on other markets appear, then with some degree of confidence, we could generalize our findings and state that changes in the market architecture have indeed a significant effect on the dynamics of wholesale price.

A drawback of using dummy variables for the reforms is that the response of the market's agents to these stimuli is not smoothed but abrupt like a step function. Furthermore, if there is a time lapse between the announcement of a reform and its implementation, this allows agents in advance, to take advantage of the reforms by adjusting themselves according to their position in the market.

We have not consider in this work, all possible interactions between different regulatory reforms. A more detail analysis is needed in future work as suggested by a growing, recent and focused on this topic, literature (child et.al, 2008).

Any required regulatory reforms that have to be applied, must be designed in such a way that facilitate the gradual development of competitive environment, allowing at the same time the minimization of the State's intervention (in areas that this is necessary) to the market as well as of the total cost of market transactions. The reforms also must create a secure environment for all participants. For this reason, any reforms that have a transitional characteristic (as most of the RMR considered in this study) must be evaluated according to the **holistic or systemic effects** on the market due to their **interdependence** and the fact that they are inherently coupled with the fundamental variables of the market, as the modeling results described in this paper has revealed. The ultimate target, therefore, of any regulatory reform is to create motives and signals to the agents of the market, so they have a collective movement to more competitive behaviors and attitudes and suitable investment decisions. Towards this end, this paper is believed to contribute to the current efforts of the Regulator and other relevant market bodies, towards restructuring the market as well as in helping developing tools / processes / rules etc., that will allow the fair pricing of product and service available in the Greek Electricity market as well as in attracting investors in the market.

We also hope that the modeling results presented here can enhance the efforts of the reformers of the GEM by helping them in detecting more accurately the factors of uncertainty, both domestic and international, that influence heavily the Greek market and must be taken into account when designing regulatory measures to face the continuously changing internal and external environment.

However, we have to point out here that the phase out of any transitional mechanism without first a detailed and in depth examination of the side effects this might have on the current stability of the market, due to the strong interlinks and coupling nature these mechanisms have with fundamental variables and have created all these years the current market microstructure, may have an adverse impact on the available long term generation capacity. Care must be taken therefore in securing future adequate availability by avoiding a forced canceling of mechanism ready in place that may put in danger the current equilibrium in the market.

This means that the impact of the change in the method of calculating SMP in the Greek market was extremely strong. It has completely change the structure of the underlying dynamic-stochastic process of SMP i.e. its statistical properties or stylized facts, causing forcedly its future dynamics evolution to be more volatile therefore more unpredictable. This **regime switching** of the process from a stable or tranquil state to a new turbulent and unstable one, give us an example of how an external perturbation (here a change in the regulatory framework) to the stochastic system that describes SMP, no matter how small or large it is, can cause the system to behave completely differently. This is a manifestation of nonlinearity and sensitive dependence on initial conditions of the electricity market systems which are very complex and nonlinear, requiring an extensive amount of simulations before attempting any change on their structure.

Since an electricity market is a high dimensional setup (a high dimensional manifold on which SMP is evolving, as referred in the jargon of Stochastic and Dynamical System theory) it is natural to consider that all the distortions in the market as well as the history of external – "exogenous" interventions in the market, is well incorporated in the dynamics of SMP, since this variable is coupled, in a nonlinear – complex way, with the driving fundamental variables in the market. Therefore, any movements in the explanatory variables used in modeling SMP (including the regulatory reforms entered in the model as dummy variables), are well "encapsulated" in the dynamics of SMP (Bask M., et. al, 2007, Broomhead D.S. et al, 1986, Takens F., 1987, Serletis A., 2007). We have extracted both qualitative and quantitative information from the SMP time series, that reveal how the factors have shaped its behavior and at the same time, have generated as bi-products, the market's structural frictions or distortions.

Even though there is not an easy job to attribute to each separate market distortion a factor or a combination of factors, due to the fact that, as we have said the system under examination is high-dimensional, non-linear and (as a consequence) very complex, we consider that the detected stylized facts of SMP, in our present analysis, can reveal an insight of the Market's distortions. Therefore, another question that is arising is whether the measures or policies of the regulator mentioned above are the appropriate ones in a sense that they are capable of removing permanently these distortions or of removing just the symptoms.

The combined effects on SMP of low demand (as in Greece due to economic crisis) and increase generation by IPP´s were more intense in 2011 when two new CCGT plants were added and the load reached very low levels. The effects on SMP were more apparent when IPP´s underwent maintenance for some period (therefore Unit availability reduced) exerting an upward movement on SMP.

As we have mentioned earlier a crucial prerequisite for GEM to be more efficient in the link between **wholesale** (SMP) and **retail prices**. Therefore regulated prices must gradually be adjusted, in order to eliminate cross-subsidy distortions and eventually produce retail prices that reflect as close as possible wholesale market costs.

**Disclaimer**
The material in this work is for information, education, research and academic purposes only. Any opinions, proposals and positions expressed in this paper are exclusively of the authors and do not necessarily represent the views of ADMIE and CRANS, partially or unilaterally.

**Acknowledgements**

The Authors acknowledge the support of Mr. Evangelos Georgis of ADMIE S.A for his great contribution to the paper, by providing to the authors with the raw and preprocessed data and by participating in numerous discussions with the first of the authors on the intermediate and final results. Thanks also to Mr. P Asslanis and A. Tsalpatouros, both of ADMIE S.A, for their help in providing part of the data sets. Special thanks are given to Mrs. K. Armeni of ADMIE S.A for the typing and general care of this manuscript.